\documentclass[manuscript]{aastex63}

\usepackage{gensymb}
\usepackage[normalem]{ulem}
\usepackage{hyperref}
\setcitestyle{super}

\shorttitle{Lensed Star}
\shortauthors{Welch et al.}

\begin{document}

\title{A Highly Magnified Star at Redshift 6.2}


\author[0000-0003-1815-0114]{Brian Welch}
\affiliation{Center for Astrophysical Sciences, Department of Physics and Astronomy, The Johns Hopkins University, 
3400 N Charles St. 
Baltimore, MD 21218, USA}

\author[0000-0001-7410-7669]{Dan Coe}
\affiliation{Space Telescope Science Institute (STScI), 
3700 San Martin Drive, 
Baltimore, MD 21218, USA}
\affiliation{Association of Universities for Research in Astronomy (AURA) for the European Space Agency (ESA), STScI, Baltimore, MD, USA}

\author[0000-0001-9065-3926]{Jose M. Diego}
\affiliation{Instituto de F\'isica de Cantabria (CSIC-UC). Avda. Los Castros s/n. 39005 Santander, Spain}

\author[0000-0002-0350-4488]{Adi Zitrin}
\affiliation{Physics Department,
Ben-Gurion University of the Negev, P.O. Box 653,
Be'er-Sheva 84105, Israel}

\author[0000-0003-1096-2636]{Erik Zackrisson}
\affiliation{Observational Astrophysics, Department of Physics and Astronomy, Uppsala University, Box 516, SE-751 20 Uppsala, Sweden}

\author[0000-0001-7399-2854]{Paola Dimauro}
\affiliation{Observatório Nacional, Ministério da Ciencia, Tecnologia, Inovaçãoe Comunicações, Rua General José Cristino, 77, São Cristóvão,20921-400, Rio de Janeiro, Brazil}

\author[0000-0002-6090-2853]{Yolanda Jim\'enez-Teja}
\affiliation{Instituto de Astrof\'isica de Andaluc\'ia, Glorieta de la Astronom\'ia s/n, 18008 Granada, Spain}

\author[0000-0003-3142-997X]{Patrick Kelly}
\affiliation{School of Physics and Astronomy, University of Minnesota, 116 Church Street SE, Minneapolis, MN 55455, USA}

\author[0000-0003-3266-2001]{Guillaume Mahler}
\affiliation{Department of Astronomy, University of Michigan, 1085 S. University Ave, Ann Arbor, MI 48109, USA}
\affiliation{Institute for Computational Cosmology, Durham University, South Road, Durham DH1 3LE, UK}
\affiliation{Centre for Extragalactic Astronomy, Durham University, South Road, Durham DH1 3LE, UK}

\author[0000-0003-3484-399X]{Masamune Oguri}
\affiliation{Research Center for the Early Universe, University of Tokyo, Tokyo, 113-0033, Japan}
\affiliation{Department of Physics, University of Tokyo, Tokyo 113-0033, Japan}
\affiliation{Kavli Institute for the Physics and Mathematics of the Universe (Kavli IPMU, WPI), University of Tokyo, Chiba 277-8582, Japan}

\author[0000-0002-0474-159X]{F.X.~Timmes}
\affiliation{School of Earth and Space Exploration, Arizona State University, Tempe, AZ 85287, USA}
\affiliation{Joint Institute for Nuclear Astrophysics - Center for the Evolution of the Elements, Tempe, AZ 85287, USA}

\author[0000-0001-8156-6281]{Rogier Windhorst}
\affiliation{School of Earth and Space Exploration, Arizona State University, Tempe, AZ 85287, USA}

\author[0000-0001-5097-6755]{Michael Florian}
\affiliation{Department of Astronomy, Steward Observatory, University of Arizona, 933 North Cherry Avenue, Tucson, AZ 85721, USA}

\author[0000-0001-9336-2825]{S.~E.~de~Mink}
\affiliation{Max-Planck-Institut für Astrophysik, Karl-Schwarzschild-Straße 1, 85741 Garching, Germany}
\affiliation{Anton Pannekoek Institute for Astronomy and GRAPPA, University of Amsterdam, NL-1090 GE Amsterdam, The Netherlands}
\affiliation{Center for Astrophysics \text{\textbar} Harvard \& Smithsonian, 60 Garden Street, Cambridge, MA 02138, USA}

\author[0000-0001-9364-5577]{Roberto J. Avila}
\affiliation{Space Telescope Science Institute (STScI), 
3700 San Martin Drive, 
Baltimore, MD 21218, USA}

\author{Jay Anderson}
\affiliation{Space Telescope Science Institute (STScI), 
3700 San Martin Drive, 
Baltimore, MD 21218, USA}


\author[0000-0002-7908-9284]{Larry Bradley}
\affiliation{Space Telescope Science Institute (STScI), 
3700 San Martin Drive, 
Baltimore, MD 21218, USA}

\author[0000-0002-7559-0864]{Keren Sharon}
\affiliation{Department of Astronomy, University of Michigan, 1085 S. University Ave, Ann Arbor, MI 48109, USA}

\author[0000-0002-4853-1076]{Anton Vikaeus}
\affiliation{Observational Astrophysics, Department of Physics and Astronomy, Uppsala University, Box 516, SE-751 20 Uppsala, Sweden}
                   
\author[0000-0003-0503-4667]{Stephan McCandliss}
\affiliation{Center for Astrophysical Sciences, Department of Physics and Astronomy, The Johns Hopkins University, 
3400 N Charles St. 
Baltimore, MD 21218, USA}


\author[0000-0001-5984-0395]{Maru{\v s}a Brada{\v c}}
\affiliation{Department of Physics, University of California, Davis, CA 95616, USA}

\author[0000-0002-7627-6551]{Jane Rigby}
\affiliation{Observational Cosmology Lab, NASA Goddard Space Flight Center, Greenbelt, MD 20771, USA}

\author[0000-0003-1625-8009]{Brenda Frye}
\affiliation{Department of Astronomy, Steward Observatory, University of Arizona, 933 North Cherry Avenue, Tucson, AZ 85721, USA}

\author[0000-0003-3631-7176]{Sune Toft}
\affiliation{Cosmic Dawn Center (DAWN), Copenhagen, Denmark}
\affiliation{Niels Bohr Institute, University of Copenhagen, Jagtvej 128, Copenhagen, Denmark}

\author[0000-0002-6338-7295]{Victoria Strait}
\affiliation{Department of Physics, University of California, Davis, CA 95616, USA}

\author[0000-0001-9391-305X]{Michele Trenti}
\affiliation{School of Physics, University of Melbourne, Parkville VIC 3010, Australia}
\affiliation{ARC Centre of Excellence for All-Sky Astrophysics in 3 Dimensions, University of Melbourne, Parkville VIC 3010, Australia}

\author[0000-0001-9851-8753]{Soniya Sharma}
\affiliation{Observational Cosmology Lab, NASA Goddard Space Flight Center, Greenbelt, MD 20771, USA}

\author[0000-0002-8144-9285]{Felipe Andrade-Santos}
\affiliation{Clay Center Observatory, Dexter Southfield, 20 Newton Street, Brookline, MA 02445, USA}
\affiliation{Center for Astrophysics \text{\textbar} Harvard \& Smithsonian, 60 Garden Street, Cambridge, MA 02138, USA}

\author[0000-0002-8785-8979]{Tom Broadhurst}
\affiliation{Department of Theoretical Physics, University of the Basque Country UPV/EHU, Bilbao, Spain}
\affiliation{Donostia International Physics Center (DIPC), 20018 Donostia,Spain}
\affiliation{IKERBASQUE, Basque Foundation for Science, Bilbao, Spain}

\begin{abstract}

{\bf
Galaxy clusters magnify background objects through strong gravitational lensing. 
Typical magnifications for lensed galaxies are factors of a few 
but can also be as high as tens or hundreds, stretching galaxies into giant arcs
\cite{RiveraThorsen17_sunburst, Johnson17L}.
Individual stars can attain even higher magnifications given fortuitous alignment with the lensing cluster.
Recently, several individual stars at redshift {\boldmath$z \sim 1 - 1.5$} have been discovered,
magnified by factors of thousands, temporarily boosted by microlensing 
\cite{Kelly18, Rodney18, Chen19, Kaurov19_lensedstar}. 
Here we report observations of a more distant and persistent
magnified star at redshift {\boldmath$z_{\rm phot} = 6.2 \pm 0.1$},
900 Myr after the Big Bang.
This star is magnified by a factor of thousands
by the foreground galaxy cluster lens WHL0137--08\ ({\boldmath$z = 0.566$}), 
as estimated by four independent lens models.
Unlike previous lensed stars, the magnification and observed brightness (AB mag 27.2)
have remained roughly constant over 3.5 years of imaging and follow-up.
The delensed absolute UV magnitude {\boldmath$M_{UV} = -10 \pm 2$} 
is consistent with a star of mass {\boldmath$M > 50 M_{\odot}$}.
Confirmation and spectral classification are forthcoming from approved observations with the \textbf{\textit{James Webb Space Telescope}}.
}
\end{abstract}

 \section{A Single Star in the First Billion Years}
 
 The Reionization Lensing Cluster Survey (RELICS \cite{Coe19_relics}) Hubble Space Telescope (\textit{HST}) Treasury Program obtained {\it HST } Advanced Camera for Surveys (ACS) optical imaging and Wide Field Camera 3 infrared (WFC3/IR) imaging of a total of 41 lensing clusters. 
 Included in these observations was a 15\arcsec--long lensed arc of a galaxy at $z_{\rm phot} = 6.2 \pm 0.1$ \cite{Salmon2020}, designated WHL0137-zD1 and nicknamed the ``Sunrise Arc" (see Extended Data Table \ref{tab:HST} and Extended Data Figure \ref{fig:photometry} for photometry and redshift estimate details).
 Its length rivals the ``Sunburst Arc'' at $z = 2.4$, the brightest strongly lensed galaxy known
 \cite{RiveraThorsen17_sunburst,RivThor19_sunburst2}. 
 Within this $z>6$ galaxy, we have identified a highly magnified star sitting atop the lensing critical curve
 at RA, Dec = 01:37:23.232, --8:27:52.20 (J2000). 
 This object is designated WHL0137-LS, and we nickname the star ``Earendel" from the Old English word meaning ``morning star", or ``rising light". 
 Follow-up {\it HST } imaging revealed Earendel is not a transient caustic crossing event;
 its high magnification has persisted for 3.5 years (see Extended Data Figure \ref{fig:variation}).

\begin{figure}[p!]
     \centering
     \includegraphics[width=0.8\textwidth]{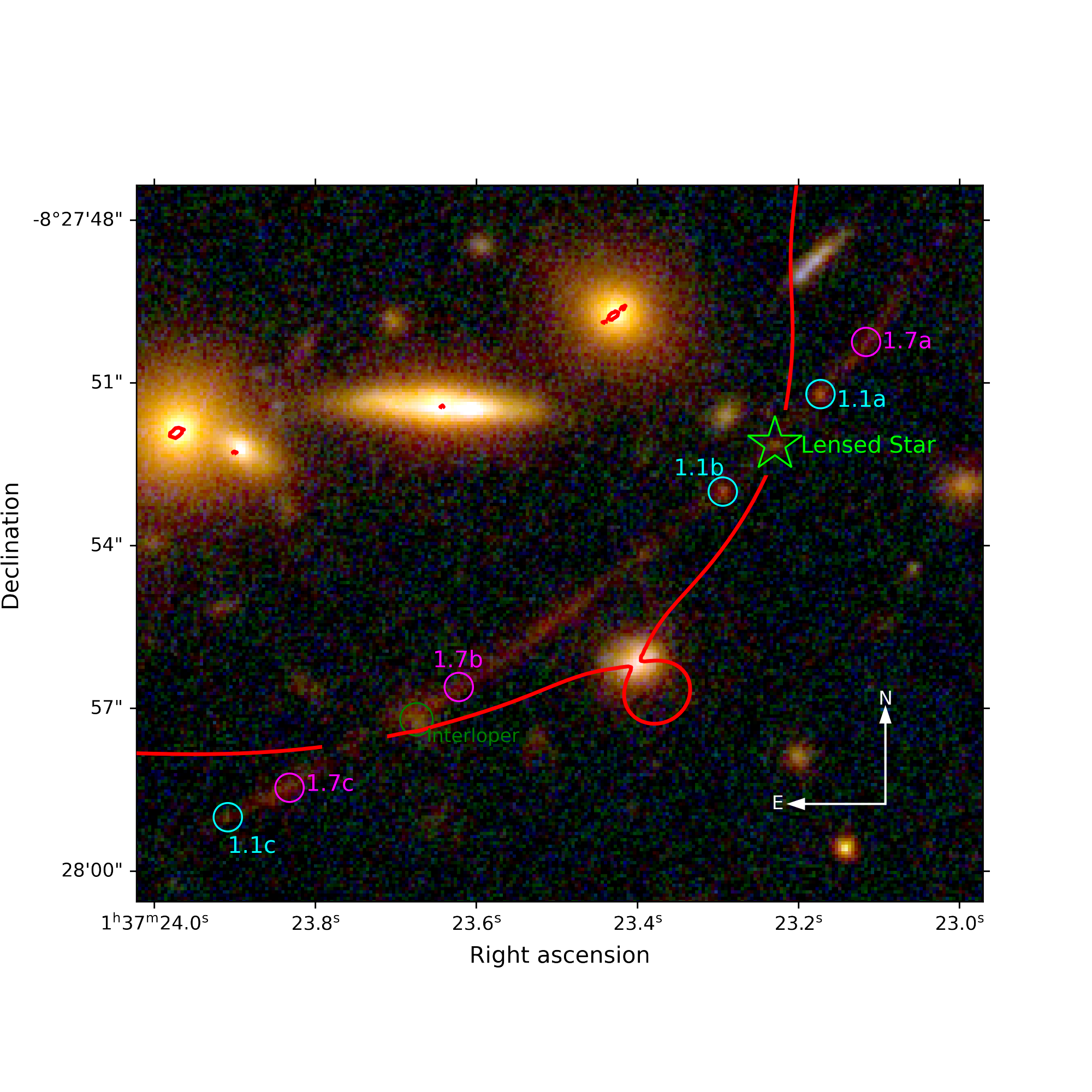}
     \caption{{\bf Labeled color image of WHL0137-zD1}
     The Sunrise Arc at $z = 6.2$ is the longest lensed arc of a galaxy at $z > 6$, with an angular size on the sky exceeding 15 arcseconds. The arc is triply-imaged and contains a total of seven star-forming clumps; the two systems used in lens modeling are circled, with system 1.1 in cyan and system 1.7 in magenta. The highly magnified star Earendel is labeled in green. The best-fit lensing cluster critical curve from the Light-Traces-Mass (LTM) model is shown in red, broken where it crosses the arc for clarity. The color composite image shows the F435W filter image in blue, F606W + F814W in green, and the full WFC3/IR stack (F105W + F110W + F125W + F140W + F160W) in red. }
     \label{fig:arclabel}
 \end{figure}

We can deduce qualitatively that the magnification of this object must be high given its position within the arc. Multiple images of lensed objects appear on opposite sides of the lensing critical curve, with the critical curve bisecting the two images. Earendel appears at the midpoint between two images of a star cluster (1.1a/1.1b in Figure \ref{fig:arclabel}). We only see one image of Earendel, indicating that its two lensed images are unresolved. Thus, the critical curve must fall near the image of the star, indicating that it will have a high magnification. 
 
Our detailed lens modeling supports this interpretation.
We model the cluster using four independent techniques:
Light-Traces-Mass (LTM \cite{Zitrin15, Zitrin09, Broadhurst05}), 
Lenstool \cite{JulloLenstool09,JulloLenstool07}, 
Glafic \cite{Oguri2010}, 
and WSLAP+ \cite{Diego07wslap2,Diego05wslap}.
To constrain these models, 
 we identify two triply-lensed clumps 1.1a/b/c and 1.7a/b/c within the Sunrise Arc
 and one triply-lensed clump within a $z \sim 3$ galaxy to the north (see Extended Data Figure \ref{fig:cluster}).
 Given these modest constraints, our lens models retain a significant degree of freedom.
 Yet all our models put Earendel within $D_{\rm crit} < 0.1''$ of the critical curve.
 
\begin{figure}[p!]
     \centering
     \includegraphics[width=0.9\textwidth]{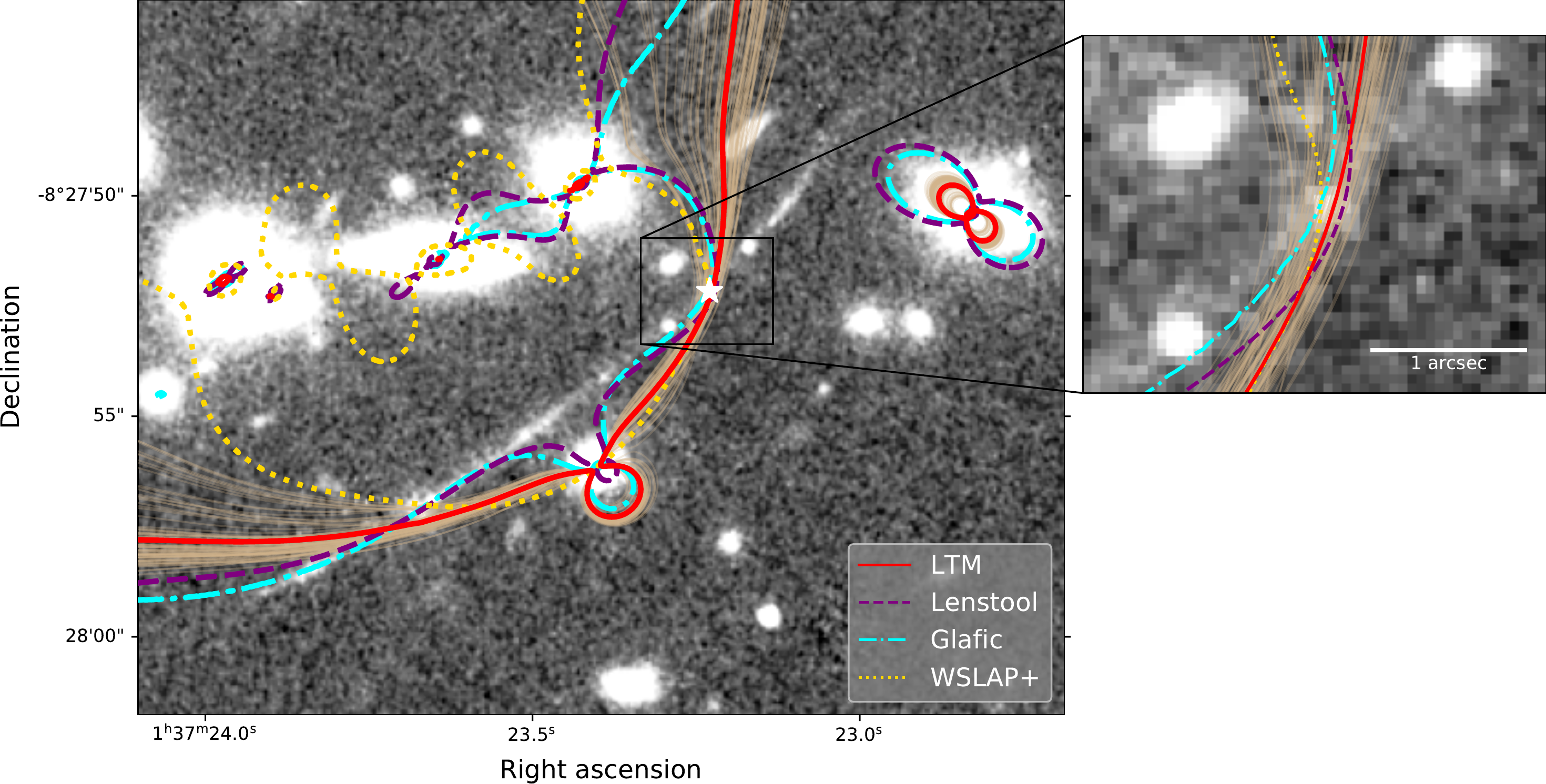}
     \caption{{\bf Strong lensing critical curves}
     Our best-fit lens models all produce critical curves that cross the lensed star Earendel within 0.1\arcsec. Additionally, 100 iterations of our LTM model drawn from the MCMC (thin tan lines) are similarly consistent, albeit with greater variance, all crossing the arc within 0.4\arcsec\ of the lensed star. Critical curves are shown for LTM (red solid), Lenstool (purple dashed), Glafic (cyan dash-dot), and WSLAP+ (yellow dotted).}
     \label{fig:modelcomp}
 \end{figure}

 Each lens model includes some uncertainty on the model parameters. To understand the effect of these uncertainties, we sampled the posterior distribution generated by the LTM lens model, and generated critical curves from each resultant parameter set. These critical curves are shown in Figure \ref{fig:modelcomp}. We find the star to be within 0.1\arcsec\ in $\sim 80 \%$ of models, while the maximum distance reaches 0.4\arcsec. 
 
 We then derived tighter constraints on the distance to the critical curve
 based on the fact that we observe only a single unresolved object.
 If Earendel were farther from the critical curve, 
 we would see two multiple images, as with clumps 1.1a and 1.1b.
 The single image means either that its two images are spatially unresolved
 or that microlensing is suppressing the flux of one image.
 We deem the latter unlikely given 
 that this cluster has an optically thick microlensing network at the location of the star (see \S\ref{sec:microlens_method} for details). In this configuration, there are no pockets of low magnification which could hide one of the images, as the microcaustics all overlap.
 Therefore we conclude the two lensed images are unresolved 
 in the current HST WFC3/IR imaging.
 This is consistent with our original lens model-independent interpretation
 suggesting it is directly on the critical curve.
 
 We then use the fact that the two images of the star are unresolved to determine the maximum allowed distance to the critical curve. 
 We analyze super-sampled drizzled images
 and find that two lensed images would be spatially resolved if they were separated by $0.11''$ along the arc (see Extended Data Figure \ref{fig:constraints}).
 Moving each image $0.055''$ along the arc puts them $< 0.036''$ from the critical curve,
 given the angle between the arc and critical curve in the various lens models
 (see \S\ref{sec:magnification_methods} for details).
 Maximum distances to the critical curve ($D_{\rm crit}$) for each lens model are tabulated in Table \ref{tab:modelresults}.
 This is a more precise determination than is possible with the weakly constrained lens models alone. 
 
 
 Using the maximum allowed separation, we can calculate the minimum magnification of the lensed star.
 In the vicinity of the critical curve, magnifications are inversely proportional to the distance to the critical curve: 
 \begin{equation} \label{eq:mu0}
     \mu = \mu_0 / D_{\rm crit}
 \end{equation}
 where $D_{\rm crit}$ is expressed in arcseconds, and $\mu_0$ is a constant that varies between lens models \cite{Diego19}.
 The value of $\mu_0$ depends on the slope of the lens potential, with flatter potentials yielding higher values of $\mu_0$ and thus higher magnifications for a given distance (i.e. LTM), while steeper potentials give lower magnifications (i.e. Lenstool). 
 Due to the paucity of lensing constraints, we can only determine the slope of the potential to within a factor of 6. 
 However, using multiple lens models, including two Glafic models with one flatter (lower concentration $c = 1$) and one steeper ($c = 7$) potential, we are able to cover the full range of possible outcomes.

 Based on this analysis, we derive magnification estimates summarized in Table \ref{tab:modelresults}.
 Note the magnification calculated from Equation \ref{eq:mu0} accounts for only one of the two unresolved images.
 We therefore double this value to get the total magnification from the source to the unresolved image.
 At the nominal estimated distances $D_{\rm crit}$, 
 the magnification estimates range from $2\mu \sim 1400$ (Lenstool) to $\sim$8400 (LTM).
 Given the uncertainty on $D_{\rm crit}$, these values may be 0.7 -- 5.0 times smaller or larger (68\% confidence).
 Thus the full range of likely magnifications is $2\mu = 1000$ -- 40000. 
 This factor of 40 uncertainty is much larger than found for lensed galaxies with typical magnifications of a few\cite{Meneghetti17}, due to the rapid changes in magnification that occur in the vicinity of lensing critical curves. Future observations will significantly shrink these error bars.

 Lower magnifications (at larger $D_{\rm crit}$) are excluded because Earendel is unresolved.
 Higher magnifications are allowed as the star approaches the caustic. 
 However, based on the cluster stellar mass density in the vicinity of the arc (Extended Data Figure \ref{fig:sig_star_region}, Extended Data Table \ref{tab:icl}, \S\ref{sec:icl_methods}) we find that the network of microlensing caustics in the star's vicinity is optically thick (see \S\ref{sec:magnification_methods}). 
 Given this microlensing configuration we estimate the maximum magnification to be of order $\mu \lesssim 10^5$, even for a transient caustic crossing \cite{Venumadhav2017,Diego18,Diego19,Dai2021}. Microlensing also has the effect of causing fluctuations in observed brightness as the lensed star traverses the microcaustic network. However, due to the optically thick microlensing network we find a 65\% probability of the observed brightness staying within a factor of 2 over the 3.5 year span of our observations (Extended Data Figure \ref{fig:microlensing}). This is consistent with our observed factor $\sim 1.4$ variation (see \S\ref{sec:microlens_method} for details).

 Our strongest evidence that Earendel is an individual star or binary rather than a star cluster
 comes from our derived $1\sigma$ upper limit on its radius.
 This limit ranges from $r < 0.09$ pc to $r < 0.36$ pc, depending on the lens model.
 We derive these limits by comparing sheared Gaussian images of various widths to the super-sampled images
 to determine what sizes are consistent with our observations of a spatially unresolved object (\S\ref{sec:magnification_methods}).

 All of the lens models yield a radius limit that is smaller than any known star cluster, indicating that this object is more likely an individual star system. 
 The smallest compact star clusters known have typical radii of order $\sim 1$ pc, with the smallest single example known having a virial radius of $0.7$ pc \cite{Portegies-zwart2010_YMCs, arches_figer99}. 
 Our largest radius constraint is a factor of two smaller than this star cluster, while our tightest constraint is nearly an order of magnitude smaller. 
 Objects at high redshift may differ from those seen in the local universe, so we also consider observations and simulations of other high-redshift objects. Bouwens et al.\cite{Bouwens17} have measured radii as small as tens of pc for very low luminosity galaxies at $6 < z < 8$, and Vanzella et al.\cite{Vanzella19_13pc} report $r < 13$ pc star clusters in a a $z = 6.143$ galaxy that is strongly lensed though not on the lensing critical curve. Recent simulations\cite{Behrendt19_clumps} resolve star-forming clumps on scales of tens of pc in $z \sim 2$ disks. Our constraints of $r < 0.09 - 0.36$ pc probe significantly smaller scales than these state-of-the-art high-redshift studies.
 We expect future spectroscopic observations with JWST 
 to conclusively determine that Earendel is one or more individual stars rather than a star cluster.
 
 Most stars of mass $M > 15 M_{\odot}$ are observed in binary systems, with a companion at a separation of $< 10$ AU \cite{Sana12,Sana14}. This is well within our observational radius constraint, suggesting that Earendel is likely composed of multiple stars. However, the mass ratio of these binaries is generally small, $\sim 0.5$ or less \cite{Moe17}. In such systems, the light from the more massive (and thus brighter) star would dominate our observation. For our primary analysis, we therefore assume that most of the light we observe is coming from a single star. The binary case is discussed further in \S \ref{sec:stellar}.

\begin{figure*}[p!]
     \centering
     \includegraphics[width=0.9\textwidth]{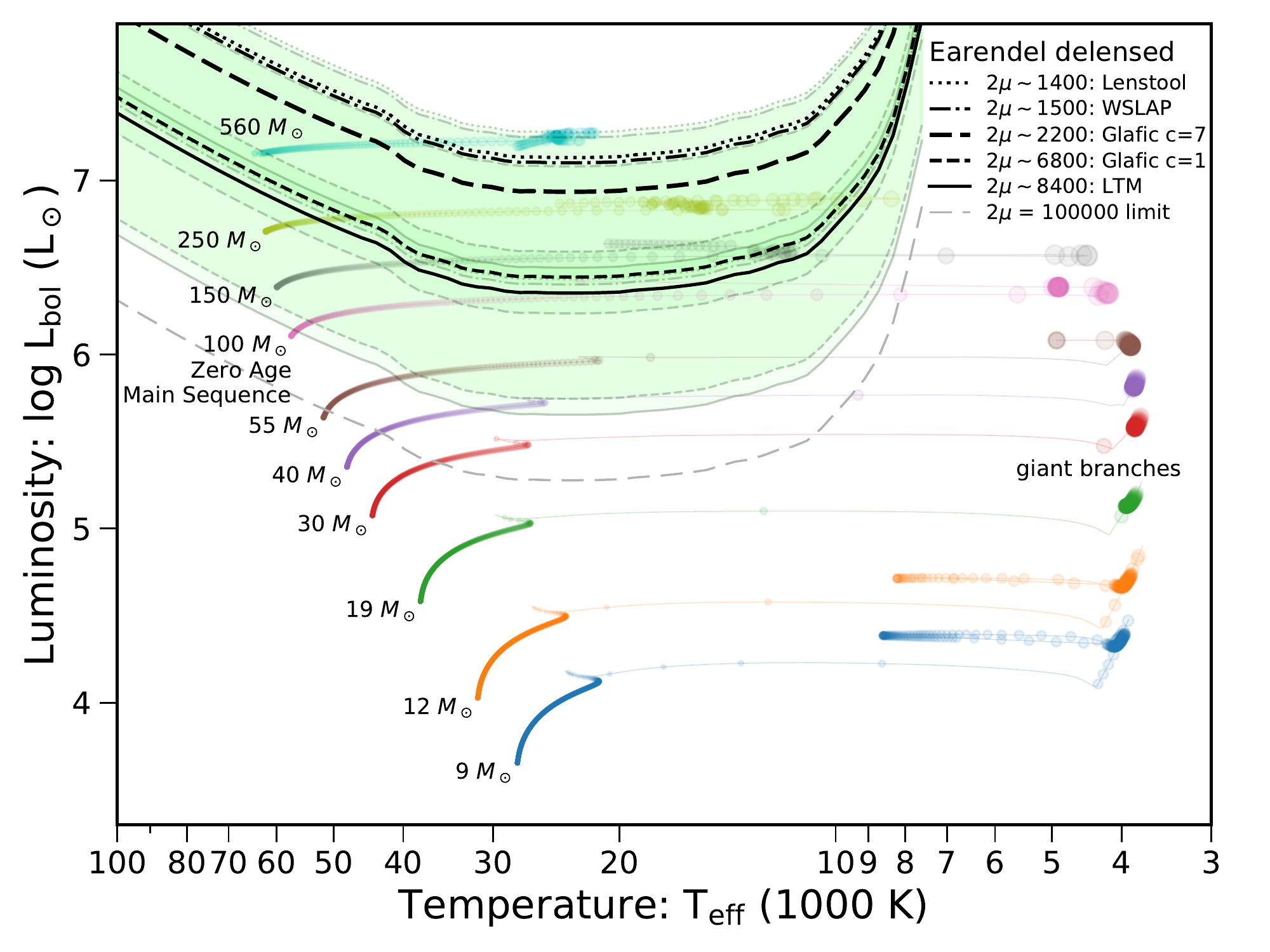}
     \caption{{\bf Lensed star constraints on the H-R diagram }
     Constraints on Earendel's luminosity and effective temperature from {\it HST } photometry
     and lensing magnification estimates, 
     shown on an HR diagram alongside BoOST stellar evolution models (colored tracks) for stars with low metallicity ($0.1 Z_\odot$) \cite{Szecsi20_boost}. 
     The green shaded region covers the $68\%$ confidence interval of our models.
     All five lens models are shown, including both Glafic models with concentration $c = 1$ and $c = 7$. 
     The theoretical upper limit magnification $\mu \sim 10^5$ is also shown; 
     it is similar to the 95\% limit for LTM. 
     Each stellar model evolution track shows points in time steps of 10,000 years with radii scaling with stellar radius.
     Our luminosity constraints favor a very massive star.
     }
     
     \label{fig:hr-diagram}
 \end{figure*}

 With a magnification between $\mu$ = 1000 -- 40000,
 we find Earendel has a delensed flux 1 -- 50 pJy (AB mag 38.7 -- 34.6)
 in the F110W filter (0.9 -- 1.4 $\mu$m),
 corresponding to an absolute UV ($1600\rm{\AA}$) magnitude $-8 > M_{AB} > -12$.
 Based on this, we constrain Earendel's luminosity as a function of temperature in the H-R diagram (Figure \ref{fig:hr-diagram}), using a combination of blackbody stellar spectra at high temperatures ($T_{\rm eff} > 40000$ K) and stellar atmosphere models at lower temperatures (details can be found in \S\ref{sec:stellar}).
 
 We compare these constraints to stellar evolution models from Bonn Optimized Stellar Tracks (BoOST\cite{Szecsi20_boost}).
 We find Earendel's derived luminosity is consistent with a single massive star 
 with initial mass $\sim 40$ -- $500 M_{\odot}$ at Zero Age Main Sequence (ZAMS).
 Note Figure \ref{fig:hr-diagram} shows a fiducial low metallicity ($0.1 Z_{\odot}$), as might be expected for a $z \sim 6$ galaxy \cite{Shimizu16}, but we explore other metallicities in \S\ref{sec:stellar} and Extended Data Figure \ref{fig:4metal}, finding this does not significantly change our mass estimates given the currently large uncertainties.
 This single star would either be a massive O-type star on the main sequence with effective temperature $\sim$60000 K and mass $>100 M_{\odot}$ 
 or an evolved O, B, or A-type star with mass $> 40 M_{\odot}$ and temperature anywhere from $\sim$8000 -- 60000 K.
 Folding in the times spent at Earendel's derived luminosity for each track and the greater relative abundances of less massive stars, we find masses between 50 -- 100 $M_{\odot}$ and temperatures above 20000K are most likely (see \S\ref{sec:stellar}, Extended Data Figure \ref{fig:star_time}).

 We estimate the probability of observing a star of mass $M \gtrsim 100 M_{\odot}$ in a caustic-crossing galaxy like the Sunrise Arc to be up to a few percent, 
 making this a fortunate yet reasonable discovery given tens of such galaxies have been observed
 (see \S\ref{sec:probability} for details).
 
 The spectral type, temperature, and mass of the star remain uncertain. Future spectroscopic observations with our approved JWST program (GO 2282; PI Coe) will determine these properties for Earendel and place it on the H-R diagram.

 \newpage
   \begin{table}[p!]
     \centering
     \begin{tabular}{lrccrrrcc}
        Lens Model & $\mu_0$ & $D_{\rm crit}$  & magnification & flux & mag & $M_{UV}$ & axis ratio & radius\\
                   &         & $''$ & $2\mu$  & nJy  & AB  &          &            & pc \\ 
        \hline 
LTM             & 	113 & 	0.027 & 	$8400^{+33600}_{-2400}$ & 	$6^{+2}_{-5}$ & 	$37.0^{+1.7}_{-0.4}$ & 	$-9.8^{+1.7}_{-0.4}$  & 	1500 & 	$<0.09$\\
Glafic ($c=1$)  & 	69 & 	0.020 & 	$6800^{+27100}_{-2000}$ & 	$7^{+3}_{-6}$ & 	$36.8^{+1.7}_{-0.4}$ & 	$-10.0^{+1.7}_{-0.4}$ & 	760 & 	$<0.14$\\
Glafic ($c=7$)  & 	23 & 	0.020 & 	$2200^{+9000}_{-600}$ & 	$22^{+9}_{-18}$ & 	$35.6^{+1.7}_{-0.4}$ & 	$-11.2^{+1.7}_{-0.4}$ & 	930 & 	$<0.21$\\
WSLAP           & 	28 & 	0.036 & 	$1500^{+6100}_{-500}$ & 	$32^{+13}_{-26}$ & 	$35.1^{+1.8}_{-0.3}$ & 	$-11.7^{+1.8}_{-0.3}$ & 	580 & 	$<0.33$\\
Lenstool        & 	18 & 	0.026 & 	$1400^{+5500}_{-400}$ & 	$36^{+14}_{-29}$ & 	$35.0^{+1.8}_{-0.3}$ & 	$-11.8^{+1.8}_{-0.3}$ & 	560 & 	$<0.36$\\

     \end{tabular}
     \caption{\label{tab:modelresults}
     {\bf Magnification, flux, and radius constraints across multiple lens models}
     Earendel results from each lens model: 
     magnification normalization $\mu_0$,
     nominal distance $D$ from critical curve,
     resulting magnification $2\mu$ (sum of two lensed images),
     delensed flux in HST F110W filter (0.9 -- 1.4 $\mu$m),
     delensed F110W magnitude,
     absolute UV magnitude ($1600\rm{\AA}$),
     model axis ratio of lensed image,
     radius upper limit.
     68\% confidence limit uncertainties are shown for all quantities.
     }
 \end{table}

\newpage 

 \newpage

 \newpage
 
 \clearpage
 \pagebreak

 \setcounter{figure}{0}
 \makeatletter
 \renewcommand{\fnum@figure}{{\bf Extended Data Figure \thefigure}}
 \makeatother
 
 \setcounter{table}{0}
 \makeatletter
 \renewcommand{\fnum@table}{{\bf Extended Data Table \thetable}}
 \makeatother
 
  \section{Methods}

 \subsection{Data}
 
 The galaxy cluster WHL0137$-$08\ (RA = 01:37:25, Dec = --8:27:25 [J2000])
 was originally discovered as an overdensity of red luminous galaxies in SDSS images \cite{WHL12},
 later confirmed at $z = 0.566$ \cite{WenHan15}
 based on SDSS DR12 spectroscopy \cite{SDSS3}.
 WHL0137$-$08\ was also ranked as the 31st most massive cluster 
 ($M_{500} \sim 9 \times 10^{14} M_\odot$)
 identified in the Planck all-sky survey PSZ2 catalog \cite{PSZ2}
 that detected clusters via the Sunyaev-Zel'dovich (SZ) effect
 on the CMB, or Cosmic Microwave Background \cite{SZ}.
 
 WHL0137$-$08\ was observed with {\it HST } as part of the RELICS Treasury program ({\it HST } GO 14096)\cite{Coe19_relics}. RELICS obtained shallow imaging of 41 lensing clusters, with single-orbit depth in the ACS F435W, F606W, and F814W optical filters, and a total of two orbits divided between four WFC3/IR filters (F105W, F125W, F140W, and F160W). These observations were split over two epochs separated by 40 days for most clusters, including WHL0137$-$08\ observed 2016-06-07 and 2016-07-17.
 
 Salmon et al.\cite{Salmon2020} performed a search for high-redshift galaxies within the RELICS data, and among that sample found the 15\arcsec\ long arc at $z_{\rm phot} = 6.2 \pm 0.1$, here dubbed the Sunrise Arc. This impressive arc warranted followup imaging with \textit{HST}, which was obtained 2019-11-04 and 2019-11-27 (PI Coe; {\it HST } GO--15842). This follow-up included an additional 5 orbits of ACS F814W imaging, along with 2 orbits each of ACS F475W and WFC3/IR F110W. These images were again split over two epochs, this time separated by 23 days. The final image was obtained 3.5 years after the first. These data were co-added to produce a full-depth image, while single epoch images were also produced to allow study of the variability of the star. Images were processed the same way as the original RELICS data \cite{Coe19_relics}. Total exposure times and limiting magnitudes in each bandpass for our co-added images are listed in Extended Data Table \ref{tab:HST}.
 
We note that this cluster has also been observed with Spitzer as part of the Spitzer-RELICS program (PI Brada{\v c}). An attempt was made to extract IR fluxes from these observations \cite{Strait21}, however reliable photometry could not be obtained due to blending with brighter cluster member galaxies nearby. 
  

 \subsection{Photometry, Redshift, and SED Fitting}
 
 We measured photometry using Source Extractor v2.19.5 \cite{sextractor}
 following procedures detailed in Coe et al. \cite{Coe19_relics}.
 The Sunrise Arc was detected as 18 source segments.
 We summed the flux measured in all segments, and summed the flux uncertainties in quadrature.
 Extended Data Table \ref{tab:HST} provides this total photometry for the Sunrise Arc,
 along with photometry for Earendel, which we identified as one of the 18 segments.
 
 We discarded the foreground interloper circled in Figure \ref{fig:arclabel} from our analysis,
 based on its slight positional offset, extended size, and colors consistent with a cluster member.
 Removing the interloper only increases the resulting photometric redshift by 0.1.
 
\begin{figure}[p!]
     \centering
     \includegraphics[width=0.9\textwidth]{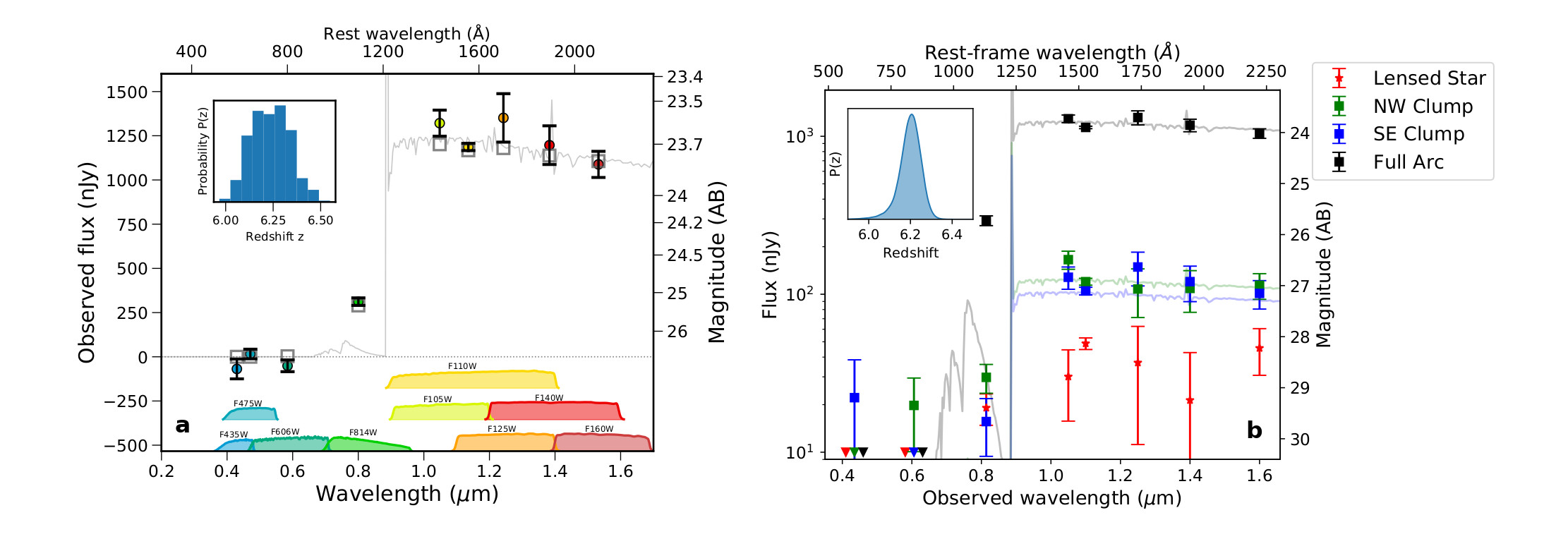}
     \caption{{\bf Photometry of the Sunrise Arc and Earendel}
     \textbf{a,} {\it HST } photometry with $1\sigma$ error bars, SED fit, and redshift probability distribution for the Sunrise Arc using the photometric fitting code BAGPIPES.
     The arc shows a clear Lyman break feature, and has a photometric redshift $z = 6.24 \pm 0.10$ (68\% CL).
     \textbf{b,} {\it HST } photometry for the full arc (black), clumps 1.1a/b (green/blue), and Earendel (red), with associated $1\sigma$ error bars.
     BPZ yields a photometric redshift of 
     $z_{\rm phot} = 6.20 \pm 0.05$ (inset; 68\% CL), 
     similar to the BAGPIPES result. 
     Clumps 1.1a/b have similar photometry, strengthening the conclusion that they are multiple images.      Note both BPZ and BAGPIPES find significant likelihood only between $5.95 < z < 6.55$ for the Sunrise Arc.
     }
     \label{fig:photometry}
 \end{figure}

 We measure the Sunrise Arc's photometric redshift using two methods:
 BPZ \cite{Benitez00bpz, Coe06bpz} and BAGPIPES \cite{Carnall18_bagpipes}.
 BPZ uses 11 spectral models (plus interpolations yielding 101 templates)
 spanning ranges of metallicities, dust extinctions, and star formation histories
 observed for the vast majority of real galaxies \cite{Coe19_relics}.
 BPZ also includes a Bayesian prior on the template and redshift given an observed magnitude in F814W.
 We allowed redshifts up to $z < 13$.
 BPZ yields a photometric redshift of $z_{\rm phot} = 6.20 \pm 0.05$ (68\% CL)
 without any significant likelihood below $z < 5.9$ (Extended Data Figure \ref{fig:photometry}).
 
 BAGPIPES generates model spectra based on physical parameters, 
 then efficiently searches a large multidimensional parameter space 
 to measure best-fitting parameters along with uncertainties.
 We ran BAGPIPES fitting simultaneously to redshift and physical parameters 
 as detailed in Strait et al.\cite{Strait21}.
 Our choices do not significantly affect the photometric redshift, 
 but we summarize them here.
 We used synthetic stellar populations from BPASS v2.2.1 \cite{Eldridge17_bpass} 
 with nebular reprocessing and emission lines added by the photoionization code CLOUDY \cite{Ferland17}.
 We used a delayed star formation history that initially rises then falls via $SFR(t) \propto t ~ e^{-t / \tau}$.
 We use the BPASS IMF ``imf135\_300'': Salpeter\cite{SalpeterIMF} slope $\alpha = -2.35$ for $0.5 < M/M_{\odot} < 300$, 
 and a shallower slope of $\alpha = -1.3$ for lower mass stars $0.1 < M/M_{\odot} < 0.5$.
 In our BAGPIPES modeling of the Sunrise Arc, 
 we left redshift as a free parameter ($z < 13$),
 along with dust (up to $A_V = 3$ mag),
 stellar mass ($10^6$ -- $10^{14} M_\odot$),
 metallicity ($0.005 - 5$ $Z_\odot$),
 ionization parameter ($2 < log(U) < 4$),
 age (from 1 Myr up to the age of the universe), 
 and SFR exponential decay time $\tau$ (100 Myr -- 10 Gyr).
 Dust extinction is implemented with the Calzetti et al.\cite{Calzetti00} law,
 and we assume twice as much dust around all HII regions in their first 10 Myr.
 %
 %
 
 The resulting best fit and redshift likelihood distribution are shown in Extended Data Figure \ref{fig:photometry}.
 BAGPIPES yields a redshift estimate $z = 6.24 \pm 0.10$, 
 similar to the BPZ result without any significant likelihood at lower redshifts.
 We tried explicitly exploring lower redshift $z < 4$ solutions,
 including old and/or dusty galaxies with intrinsically red spectra 
 that can result in photometric redshift degeneracies for some high-redshift galaxies.
 But in our case, none of those red spectra can reproduce the flat photometry observed for the Sunrise Arc
 in the near infrared (1.0 -- 1.6 $\mu$m).
 
 BAGPIPES further yields estimates of the dust $A_V = 0.15 \pm 0.10$ mag and mass-weighted age $135 \pm 60$ Myr,
 but no strong constraints on metallicity or ionization parameter.
 Any estimates of stellar mass and SFR include uncertainties due to the lensing magnification.
 But simply adopting a fiducial magnification of 300 for the full arc from the LTM model,
 we estimate a stellar mass $\sim 10^{7.5\pm 0.2} M_\odot$
 and current SFR $\sim 0.3 \pm 0.1 \, M_{\odot} \text{ yr}^{-1}$,
 not including magnification uncertainties.

 \subsection{Variability} \label{sec:variability}
 
 Microlensing simulations suggest that the magnification, and thus observed flux, of Earendel should remain relatively constant over time. However, some variation is expected as the star traverses the microlensing caustic network. A factor of 1--3 difference in observed flux would be expected given these simulations (see \S\ref{sec:microlens_method}). 
 
\begin{figure}[p!]
     \centering
     \includegraphics[width=0.9\textwidth]{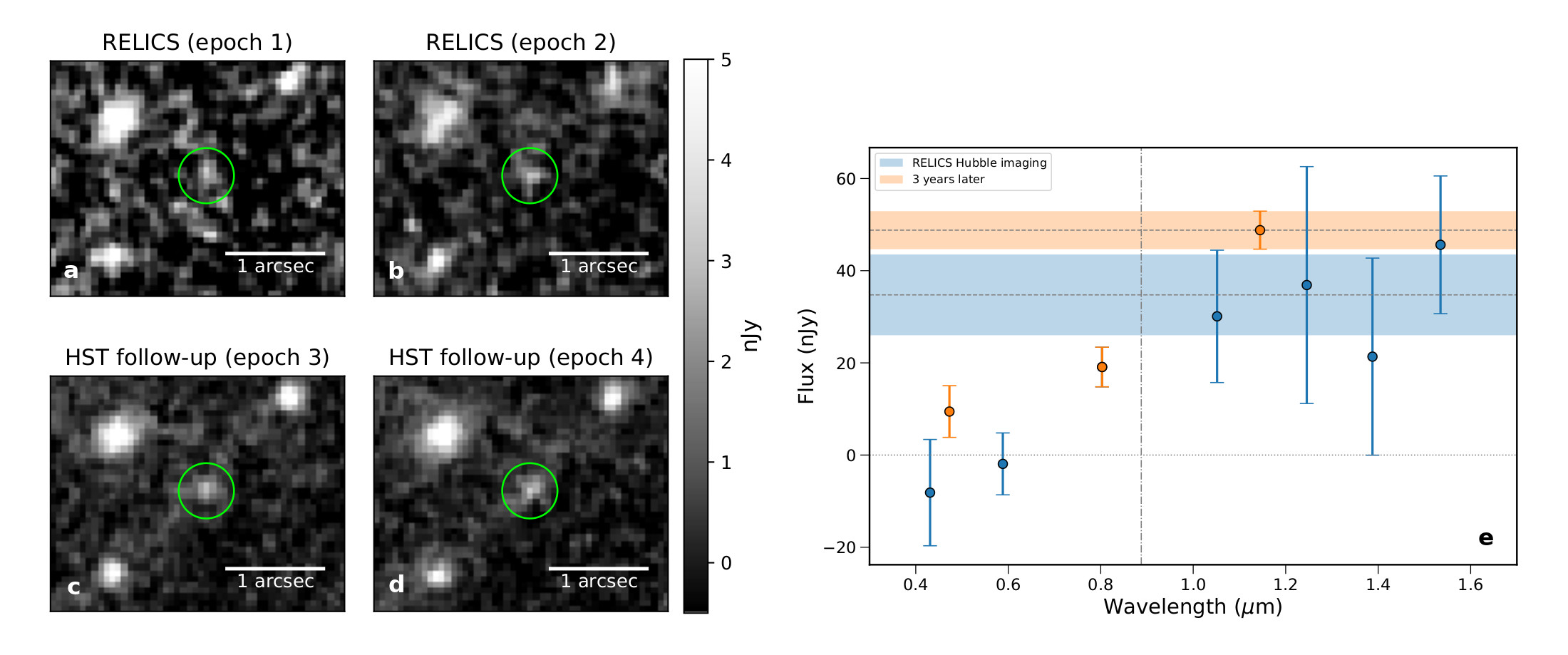}
     \caption{{\bf Lensed star variability across observations}
     Earendel has remained consistently bright across 3.5 years of {\it HST } imaging. Panels \textbf{a-d} show WFC3/IR images of the lensed star (circled in green) across four epochs. Panels \textbf{a} and \textbf{b} show epochs 1 and 2 respectively, taken as part of RELICS, and are a sum of the IR imaging in 4 filters F105W+F125W+F140W+F160W from each epoch (one orbit each). Panels \textbf{c} and \textbf{d} show follow-up F110W imaging taken in epochs 3 and 4 respectively (one orbit each, in a more efficient filter). Panel \textbf{e} shows a plot of the original RELICS photometry (blue) compared to the follow-up photometry (orange), each with $1\sigma$ error bars. The blue band is the weighted average of the original RELICS IR fluxes ($35 \pm 9$ nJy, 68\% CL), while the orange band is the new F110W flux ($49 \pm 4$ nJy, 68\% CL).}
     \label{fig:variation}
 \end{figure}

 To assess the variability of the star, we study the four available epochs of {\it HST } imaging separately. 
 Images from each epoch are shown on the left in Extended Data Figure \ref{fig:variation}, where Earendel is circled in green. 
 Each image shows 1 orbit of WFC3/IR imaging:
 for RELICS, this is a WFC3/IR stack F105W+F125W+F140W+F160W,
 while the follow-up imaging consists of F110W.
 Our analysis is complicated by the fact that RELICS and the follow-up imaging did not use the same WFC3/IR filters.  
 We measure $49 \pm 4$ nJy in the F110W imaging (sum of epochs 3 and 4)
 and derive a single value $35 \pm 9$ nJy in RELICS (epochs 1 and 2)
 from a weighted average of the fluxes measured in the four WFC3/IR filters.
 The results are shown as horizontal bands in Extended Data Figure \ref{fig:variation},
 along with the fluxes measured in each filter individually.
  Note summing only the RELICS filters F105W+F125W (closest to F110W) yields $34 \pm 15$ nJy,
  similar to the result from the full stack with larger uncertainty.

 We find that the IR flux may have varied by a factor of $\sim 1.4$ across the epochs. However, the large uncertainties on our measured flux values mean that this number is consistent with no variation. Thus we conclude that we see no significant variation across our observations. This low level of variation is consistent with our microlensing simulation results. Future observations with {\it HST } and \textit{JWST} will further explore the variability of this highly magnified object. 
 
 \subsection{Lens Modeling}
 
 The Sunrise Arc is a highly magnified system, and the lensed star Earendel was identified by the large magnification given by our best fit lens models. Strong lensing magnifications have steep gradients in the vicinity of lensing critical curves, so to evaluate the validity of our interpretation of the arc and lensed star we have taken great care in modeling the lensing cluster. We have constructed a total of five lens models using four independent modeling programs, LTM \cite{Zitrin09, Zitrin15,Broadhurst05}, Lenstool \cite{JulloLenstool07,JulloLenstool09}, Glafic \cite{Oguri2010}, and WSLAP+ \cite{Diego05wslap,Diego07wslap2}.

 We utilized a total of three sets of multiple images in our lens model optimization. The multiple image systems are highlighted in Extended Data Figure \ref{fig:cluster}, with each arc shown in detail within that figure. System 1 is the Sunrise Arc at $z_{\rm phot} = 6.2$, and system 2 consists of three images of a bright blue knot at $z_{\rm phot} = 3.1$. Within the Sunrise Arc, we use two sets of multiple images, labeled 1.1 and 1.7. Positions of multiple images 1.1 and 1.7 are defined using the F110W data, as this filter has the strongest detections for each component of the arc (SNR ranges from 7 for the faintest feature to over 20 for the brightest). 1.1 is a compact star-forming clump within the galaxy. Two of the images of 1.1 bracket the star, and are themselves highly magnified. The third image appears much fainter at the southeastern end of the arc. The apparent difference in surface brightness between these clumps is due to the fact that all three are unresolved. This fainter third image was not included in the Glafic models, while it was included in the LTM, Lenstool, and WSLAP+ models. Comparisons with LTM and Lenstool models made without including this third image show insignificant deviation from models including the third image, indicating that it may not be critical to include when our other constraints are used. The images of 1.7 consist of a clump near the opposite end of the arc. This clump is closer to the center of the host galaxy, and so it is harder to pick out among the flux of the host galaxy. However, our lens models support its positioning, and it allows us to include the full length of the arc in the lens model optimization by including a positional constraint at each end. No additional counter-images of the arc are predicted by our lens models. 
 
 Cluster member galaxies were selected via the cluster red sequence \cite{Ellis97,Stanford98}. 
 We selected galaxies along the cluster red sequence in two colors, (F435W -- F606W) and (F606W -- F814W). We also included a redshift selection, only including galaxies in the range $0.35 \leq z_{\rm phot} \leq 0.8$, bracketing the cluster redshift of $z_{cluster} = 0.566$.
 After selecting galaxies that fit these criteria, we performed a visual inspection to confirm or remove  cluster members based on morphology. 
 Finally, we chose to include two small galaxies near the Sunrise Arc which are more questionable cluster members. These small galaxies, marked as C and D in Extended Data Figure~\ref{fig:cluster}, may be small cluster members or background galaxies, and would normally have been too faint to include in our lens models. 
 They are only included here due to their proximity to the arc, and thus their increased potential to impact the lens magnification in this region. 
 Their questionable status as cluster members led us to leave them to be freely optimized in the LTM and \texttt{Lenstool} models, while galaxy C was excluded from the WSLAP+ model. 
 This galaxy is given low mass in other models (downweighted relative to its observed magnitude), so its exclusion from the WSLAP+ model causes only minor changes in the shape of the critical curve.
 
 \subsubsection{Light-Traces-Mass Lens Model}
 
 The primary lens model used for this analysis was created using the Light-Traces-Mass (LTM) method \cite{Zitrin09, Zitrin15,Broadhurst05}. As the name suggests, the LTM method assumes that light approximately traces mass within the lensing cluster. Each cluster member is assigned a power-law mass density distribution, with the overall scaling proportional to the measured flux of the galaxy. The sum of these galaxy-scale masses is then smoothed with a Gaussian kernel of variable width to represent the cluster-scale dark matter distribution. This simple description of the lens allows a rudimentary lens model to be created without multiple image constraints set, as mass is assigned based on cluster member positions and fluxes. Multiple image candidates can then be checked against the initial lens models, which in turn are iteratively refined. 
 
 In this case, we began with the a priori assumption that the Sunrise Arc consisted of two images of a single source galaxy, reflected once across the critical curve. However, initial models disfavored this interpretation, and some exploration revealed that a triply imaged galaxy was the only way to reproduce the full length of the arc. The other multiply imaged system at $z \sim 3$ was initially assumed to be triply imaged, with three clear knots showing similar photometry and morphology. These knots were confirmed by the exploratory LTM models. 
 
 After the initial explorations solidified our interpretation of the multiple image constraints, we optimized the model using the standard LTM minimization algorithm. Briefly, the distances between true multiple image locations and model-predicted positions are minimized using a $\chi ^2$ function. The minimization is done with a Monte-Carlo Markov Chain (MCMC) using a Metropolis-Hastings algorithm \cite{Hastings1970}. 
 
\begin{figure*}[p!]
     \centering
     \includegraphics[width=0.9\textwidth]{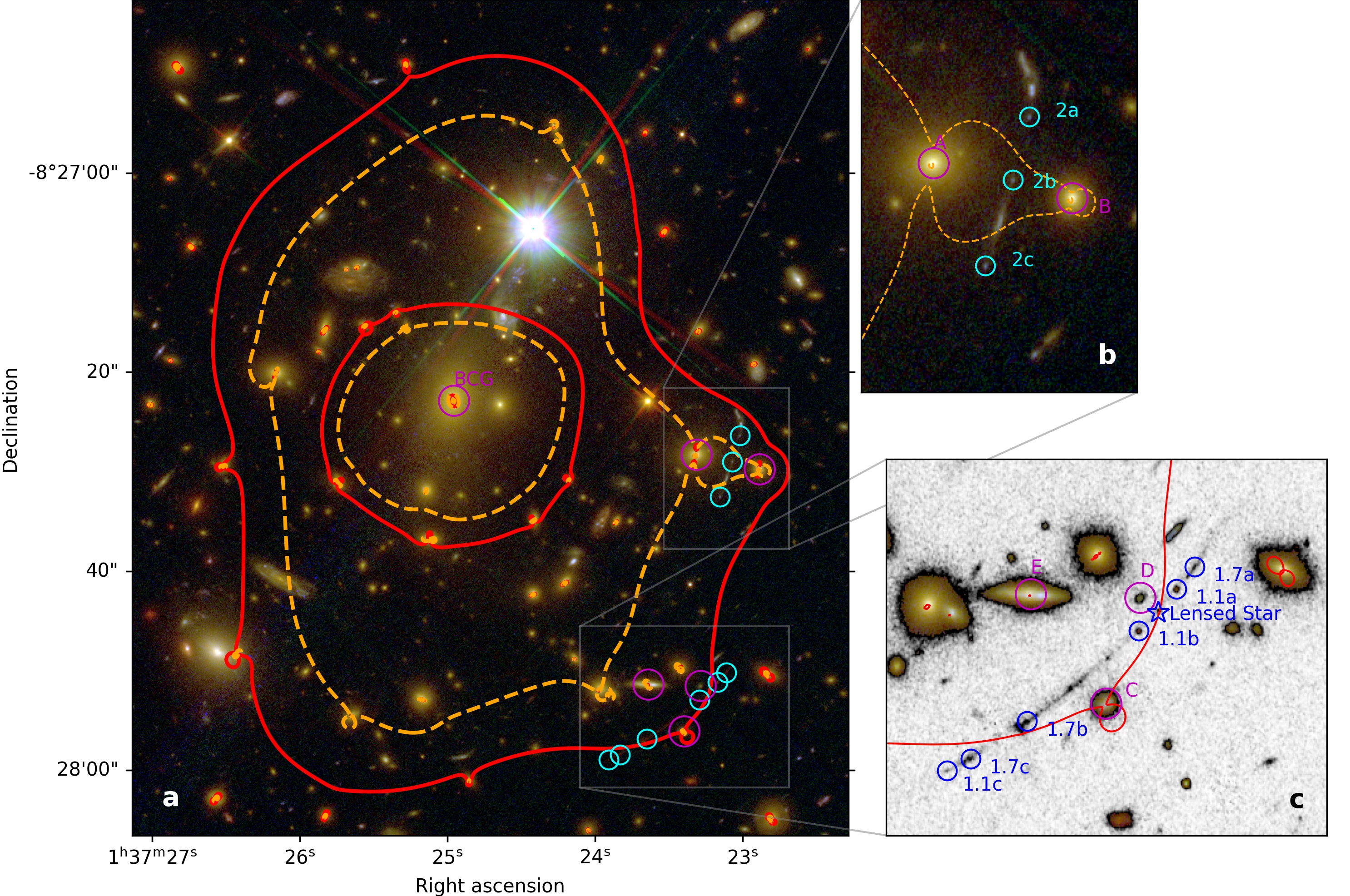}
     \caption{{\bf Strong lens modeling constraints for WHL0137$-$08}
     {\bf a,} {\it HST } composite image of WHL0137$-$08, a massive galaxy cluster at $z = 0.566$ which lenses the Sunrise Arc. Multiple images of the two lensed galaxies used in the lens modeling are marked in cyan and labeled in zoomed outsets. Cluster member galaxies circled in magenta are those freely optimized in both the LTM and Lenstool lens models. Critical curves are shown for the best-fit LTM model. The dashed orange curve is at $z = 3.1$, the same photometric redshift as multiple image system 2 (shown in {\bf b}), while the solid red curve is at $z = 6.2$, the photometric redshift of the Sunrise Arc (system 1, shown in {\bf c}). The lensed star Earendel lies directly between 1.1a and 1.1b. Note that 1.1c appears fainter than its counter-images 1.1a/b due to its lower magnification and all of these images being unresolved.}
     \label{fig:cluster}
 \end{figure*}

 During optimization, we allowed the relative weights of six galaxies (circled in Extended Data Figure \ref{fig:cluster}) to be freely optimized. This allows the model additional freedom where needed.  
 Each free galaxy is allowed to vary individually in brightness (and thus mass) by a factor ranging from 0.5 to 3 with a flat prior.
 The brightest cluster galaxy (BCG) is left free as standard. Beyond that, we allow the two bright cluster members near system 1 (labeled A and B in Extended Data Figure \ref{fig:cluster}) to vary due to their proximity to our multiple image constraint. Similarly, the relative weights of the galaxies labeled C and D were left free due to their proximity to the Sunrise Arc. Additionally, the membership status of these galaxies is uncertain, as described above. Allowing their weights to vary effectively covers the range of possibilities, from these being true cluster members to unrelated background galaxies. Finally, galaxy E appears to be a spiral disk galaxy viewed edge-on. Such galaxies follow a different mass-to-light ratio than elliptical galaxies, so allowing it to vary accounts for this difference. Relative galaxy weights in the best fit model range from 0.8 to 2.2.
 
 In addition to the multiple image constraints described above, we added flux constraints from the bright knots bracketing the star, and added parity constraints to all image systems. The flux constraint helps to pinpoint the location of the critical curve crossing, which is of critical importance to our analysis of this object. The parity constraints serve to counteract the proximity of our multiple images. Since our images are separated by as little as 1\arcsec, the MCMC optimization often found solutions with a single image appearing near the midpoint of the arc. While this does provide a low $\chi ^2$, it is clear that the relensed images do not look anything like the true arc. The parity constraint requires that the critical curves pass between multiple images, giving them opposite parity. If this constraint is not met, a penalty is added to the $\chi ^2$ function, allowing the model to avoid these local minima and find the true solution. 
 
 The LTM model provides the most accurate reconstruction of the full length of the Sunrise Arc, thus it is the one we take as our overall best fit model.

 \subsubsection{\texttt{Lenstool} Lens Model}
 
 The accuracy of the cluster lens model is of critical importance to our analysis of this object. Therefore, in order to confirm that our lensing interpretation is correct, we modeled the cluster lens using additional independent software packages. The secondary package used in this analysis was the Lenstool lens modeling software \cite{JulloLenstool07, JulloLenstool09}. 
 
 Lenstool is a parametric model that utilizes a MCMC method to sample the model parameter space. The model assigns pseudo-isothermal elliptical mass distributions (PIEMD) \cite{Limousin_piemd} to both the cluster-scale dark matter halo as well as to individual cluster member galaxies. The total mass distribution is a superposition of the cluster-scale mass distribution and the smaller galaxy-scale masses. Each PIEMD model has seven free parameters: the position ($x$, $y$), ellipticity, position angle, core radius $r_{core}$, truncation radius $r_{cut}$, and the effective velocity dispersion $\sigma_0$. Note $\sigma_0$ is not precisely the observed velocity dispersion; see Eliasd\'{o}ttir et al.\cite{Eliasdottir07} for details. 
 
 Six of the seven parameters of the PIEMD model are left free to be optimized, with the exception being the cut radius as this is not well constrained by strong lensing data alone. To keep the total number of parameters from getting too large, the parameters for the galaxy-scale masses are determined by their photometric properties, assuming a constant mass-to-light ratio. This is done using scaling relations for the velocity dispersion $\sigma_0 \propto L^{1/4}$ and truncation radius $r_{cut} \propto L^{1/2}$. The constants of proportionality are optimized freely, while the positions, ellipticities, and position angles are all fixed to what is measured photometrically using Source Extractor \cite{sextractor}. 
 
 In our modeling, we choose to leave several key galaxies free to have their velocity dispersions and radii freely optimized. These free galaxies are highlighted in Extended Data Figure \ref{fig:cluster}. As mentioned above, the spiral galaxy (E) does not follow the same M/L relation as cluster elliptical galaxies, and is therefore left free. We again leave the two small, white galaxies near the arc (C, D) free both because of their proximity to the arc, and because of their questionable status as cluster members. If these are not part of the cluster, their effect on the lensing of the arc will be much less, so we account for that by allowing their masses to vary. Finally, two galaxies are left free near the $z \sim 3$ system (A, B).
 Each free parameter is assigned a Gaussian prior centered on the parameter value given by the above scaling relations. The velocity dispersion priors are given a width of 15 km/s, and priors on radius are given a width of 5 kpc. Best fit values all fall within $\sim 2.5\sigma$ of the original value.

 \subsubsection{Glafic Lens Model}
 
 The Glafic model used here was made using the publicly available Glafic lens modeling code \cite{Oguri2010}. Glafic adopts a parametric lens modeling approach in which the lensing mass is built of multiple components, each defined by a small number of parameters (position, mass, ellipticity, and position angle). Cluster member galaxies are modeled with PIEMD mass models, while the larger cluster-scale potential is modeled with two NFW halos \cite{NFW96} placed at the positions of the brightest and second brightest cluster member galaxies. To reduce the total number of parameters in this model, the member galaxies are assumed to scale with luminosity, such that the velocity dispersion $\sigma \propto L^{1/4}$, and the truncation radius $r_{cut} \propto L^\eta$ with $\eta$ being fixed to $1$ for simplicity. The normalizations of these scaling relations are left as free parameters. The ellipticities and position angles of the member galaxies are fixed to values measured from the images using Source-Extractor \cite{sextractor}. The parameters of the lens model are optimized using a MCMC. 
 
 \subsubsection{WSLAP+ Lens Model}
 
 The final lens modeling package used in our analysis is the hybrid parametric/non-parametric WSLAP+ code \cite{Diego05wslap, Diego07wslap2}. This modeling program divides the cluster mass distribution into a compact component associated with cluster member galaxies, and a diffuse component representative of the cluster dark matter halo. The compact component assigns mass to cluster galaxies based on their luminosity via a mass-to-light ($M/L$) scaling ratio. This scaling is fit to a single value for the ensemble of cluster members during the optimization process.
 
 The diffuse mass component is defined as the superposition of Gaussians on a grid. These grid cells map the mass at any given point in the cluster, and are optimized in conjunction with the compact galaxy masses. This non-parametric aspect of WSLAP+ gives it more freedom to assign mass where a parametric model (such as LTM or Lenstool) might not. 
 
 Such non-parametric models probe a broader range of solutions, often allowing larger uncertainties for measured quantities \cite{Meneghetti17}. However, the increased freedom of this type of lens model can more readily fit atypical mass distributions, meaning it is more likely to span the true solution, particularly when the underlying mass distribution deviates from our typical mass-to-light assumptions. In this case, the increased freedom can determine if such an atypical mass distribution could explain our observations as a moderately magnified cluster of stars rather than a single star. 
 
 Instead, we find that the WSLAP+ model agrees with our parametric models, with the $z = 6.2$ critical curve crossing within 0.1\arcsec\ of Earendel in all models. This supports our interpretation of Earendel as an extremely magnified single star. 
 
 \subsection{Magnification and Size Constraints}
 \label{sec:magnification_methods}
 
\begin{figure}[p!]
     \centering
     \includegraphics[width=0.5\textwidth]{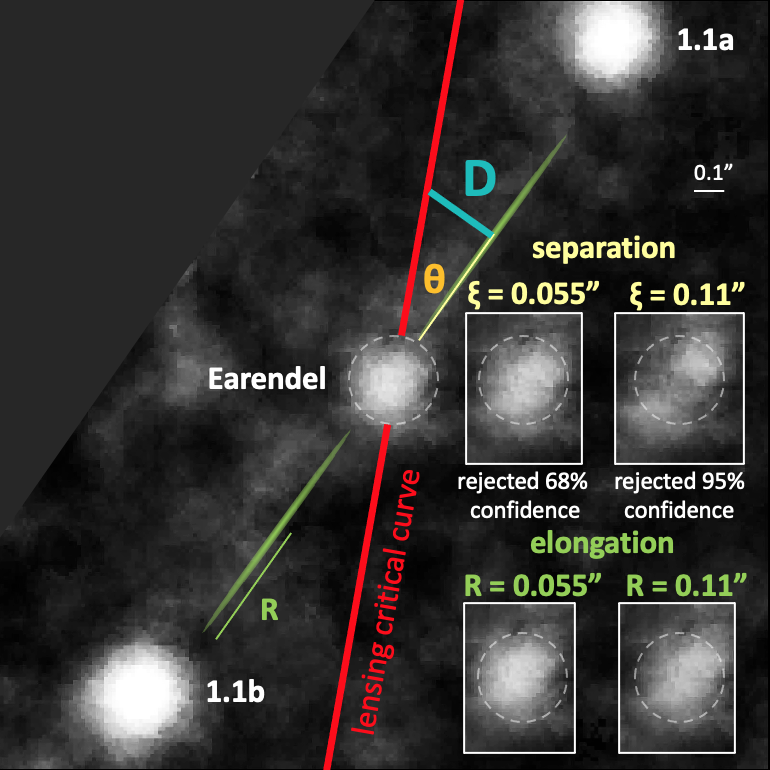}
     \caption{{\bf Size and separation upper limit measurements}
     Earendel's image is spatially unresolved.
     We manipulate this image, separating it in two or stretching it in place
     to put upper limits on its magnified radius $R < 0.055''$
     and distance $2 \xi < 0.11''$ between two unresolved images.
     These constraints allow us to calculate constraints on the intrinsic radius $r$,
     distance $D$ to the critical curve, and magnification $\mu$ for each lens model.
     Here we show a zoomed region of the arc around Earendel in a 10x super-sampled reconstruction
     of our HST WFC3/IR F110W image based on 8 drizzled exposures.
     The distances and radius labeled in the diagram are exaggerated for visibility.}
     \label{fig:constraints}
 \end{figure}

 Using the various lens models, we constrain Earendel's magnification and delensed properties as summarized in the main text.
 Based on observing a single unresolved image,
 we place upper limits on Earendel's radius and distance from the critical curve
 as illustrated in Extended Data Figure \ref{fig:constraints}.
 
 First, we model the source as a Gaussian light profile with a width $\sigma$ 
 that we refer to as the radius $r$ (e.g., 0.1 pc).
 Then we stretch this Gaussian along the arc for a given model 
 magnification $\mu = \mu_\parallel \mu_\perp$
 and axis ratio $\mu_\parallel / \mu_\perp$ where 
 the tangential magnification
 $\mu_\parallel = 1 / (1 - \kappa - \gamma)$ and
 perpendicular magnification 
 $\mu_\perp     = 1 / (1 - \kappa + \gamma)$
 for a lens model mass $\kappa$ and shear $\gamma$ at a given position (note this is normally referred to as radial magnification as defined with respect to the BCG. However that would be confusing in this context where Earendel's radius is magnified most significantly by the tangential magnification $\mu_\parallel \sim 1000$,
 and much less so by the perpendicular or radial magnification 
 $\mu_\perp \sim 1.1$ -- 2.1).
 In practice, for each model we measure $\mu_\perp$ near Earendel (it varies slowly)
 and $\mu_0$ from fitting $\mu = \mu_0 / D('')$.
 Then the observed lensed radius along the arc would be $R = \mu_\parallel r$ = $\mu_0 r / D('') \mu_\perp$.
 We convert the magnified radius to image pixels via $1'' = 5.6$ kpc (at $z = 6.2$).
 Note the resulting lensed image is approximately a 1D Gaussian line 
 (stretched almost entirely tangentially along the arc)
 convolved with the HST WFC3/IR F110W PSF (point spread function).
 
 Then by analyzing the HST images, we determine that a 1D Gaussian with a width of $\gtrsim 0.055''$ 
 would begin to appear spatially resolved.
 This width is roughly 0.4 native WFC3/IR pixels (each $0.13''$),
 or $\sim 1$ drizzled pixel ($0.06''$) in our standard data products.
 We perform this analysis on a $10\times$ super-sampled image (drizzled to $0.013''$ / pixel)
 combining the 8 F110W exposures.
 Given a model of the HST F110W PSF, we first confirm that the image of Earendel is unresolved:
 consistent with and not measurably wider than the PSF.
 Then we stretch the PSF diagonally along the arc, 
 finding that it appears unresolved when convolved with a 1D Gaussian with $\sigma = 0.055''$.
 Combining this upper limit on $R$ with the lens model estimates of $\mu_0$ and $\mu_\perp$,
 and upper limits on $D('')$ derived below,
 we determine that Earendel's intrinsic delensed radius is $r < 0.09$ -- 0.36 pc, depending on the lens model.
 
 Additionally, the radius upper limit assumes that a hypothetical star cluster would sit centered on the lensing caustic.
 This configuration would imply that our unresolved object is a merged pair of images, each showing the same half of the source cluster. 
 This specific geometry would require precise alignment of the star cluster with the lens, making it less likely. 
 A more likely geometry would be a hypothetical star cluster appearing entirely on the visible side of the caustic, now creating a merging pair of images of the full cluster. 
 This setup would decrease our radius limits by a factor of two (to $r < 0.045$ pc for LTM through $r < 0.18$ pc for Lenstool), further straining the possibility that this is a star cluster. 
 There is a possibility that we are seeing a small fraction of a larger star cluster, and that the rest of the cluster is hidden behind the caustic. 
 This is also somewhat unlikely, but we cannot rule it out.
 
 Given this radius smaller than known star clusters, 
 we determine Earendel is more likely an individual star or star system.
 Such systems are significantly smaller and would certainly appear as unresolved.
 Thus going forward, we assume Earendel's light can be modeled as a point source.
 
 Next we model Earendel as two lensed point sources separated by a distance $2\xi$.
 By analyzing the HST images, we find a similar result:
 the two images (convolved with the PSF) can be spatially resolved
 when separated by a distance $2\xi \sim 0.11''$, with $\xi \sim 0.055''$.
 These lensed images would appear separated along the arc.
 The lensing critical curve intersects the arc at an angle $\theta$
 that varies between $22\degree$ and $41\degree$, depending on the lens model.
 The distance from each lensed image to the critical curve is $D = \xi \sin \theta$.
 Thus, the maximum distance to the critical curve is $D < 0.055'' \sin \theta$,
 which varies from $D < 0.02''$ to $D < 0.036''$, depending on the lens model.

 
 Given $D$, we determine the magnification estimate $\mu$ from each model
 and constraints on radius $r$ described above, all summarized in Table \ref{tab:modelresults}.
 We note that given the strong lens model constraints (observed multiple image locations),
 the critical curve can be at any small distance $D < 0.1''$ from Earendel with roughly equal likelihood.
 This translates to a magnification likelihood $P(\mu) \propto 1 / \mu$.
 We confirm this likelihood distribution in the LTM MCMC posterior range of models.
 We note microlensing introduces additional scatter and uncertainty, but this is subdominant.
 
 Our upper limits on $D$ (68\% confidence) translate to lower limits on $\mu$, 
 given $\mu \propto 1 / D$.
 Here rather than upper limits $D < D_1$,
 we are more interested in the 68\% central confidence range, which is $0.2 D_1 < D < 1.4 D_1$ assuming a Gaussian likelihood.
 Given $\mu \propto 1 / D$ and lower limits $\mu > \mu_1$,
 the corresponding central confidence range is $0.7 \mu_1 < \mu < 5.0 \mu_1$.
 For example, at 68\% confidence LTM yields either $2\mu > 8400$, or $6000 < 2\mu < 42000$.
 This statistical uncertainty (a factor of 7) is comparable to 
 the large systematic uncertainty spanned by the various lens models (a factor of 6).
 LTM yields the highest magnification estimates, while Lenstool yields the lowest: 
 $1000 < 2\mu < 6900$ (see Table \ref{tab:modelresults}).
 Thus, rounding slightly, we quote the full uncertainty range as $1000 < 2\mu < 40000$ for Earendel's magnification.
 
 %
 
 
 We note we also attempted to measure constraints on radius and separation using a forward modeling technique as in
 Johnson et al.\cite{Johnson17model}.
 However, this method is limited to the few allowed lens models that
 each put Earendel at a discrete distance $D$ from the critical curve.
 In order to determine limits on $D$, $r$, and $\mu$, we needed to vary $D$ and $R$ smoothly.
 Forward model results for each model fell within the allowed ranges derived above.

  \subsection{Microlensing Effects}

 Previous lensed star discoveries were identified when the magnification, and thus observed brightness, temporarily increased \cite{Kelly18,Rodney18,Chen19, Kaurov19_lensedstar}. 
 These transients have relied on microlensing, wherein stars bound to the lensing cluster temporarily align with the lensed image(s) of the star, creating a brief boost to the magnification. 
 The relative transverse motions of lensing stars with respect to the lensed star
 impact the microlensing alignment and lead to the fluctuations observed in previous lensed stars. It is possible to decrease the amplitude of these microlensing fluctuations if the optical depth of microlenses increases. In situations where the magnification is extreme, microcaustics overlap in the source plane resulting in relatively small fluctuations in the flux every time a microcaustic crossing happens. The more microcaustics overlap, the less effect they have on the observed flux.  This ``more is less" microlensing effect is observed when the effective optical depth of microlensing is greater than 1 (that is, microcaustics are overlapping each other \cite{Dai2021}). In this situation, the observed flux is the sum of the fluxes from all microimages. Since the number of microimages scales with the number of overlapping microcaustics, crossing one microcaustic results in a smaller relative change in the total flux for a larger number of overlapping microcaustics \cite{Dai-Pascale2021}.

 \subsubsection{Diffuse Light Calculation} \label{sec:icl_methods}
 
 Our microlensing simulations depend on the number density of stars in the line of sight to Earendel. These can be a combination of stars or stellar remnants in the wings of cluster member galaxies, and stars or stellar remnants not bound to any galaxy that make up the ICL. To facilitate our microlensing analysis, we measured the cluster stellar mass density in the region around the Sunrise Arc, combining both the ICL and faint wings of cluster galaxies. From the stellar mass density, we can then calculate the number density of stars using an assumed initial mass function.
 
\begin{figure}[p!]
     \centering
     \includegraphics[width=0.8\textwidth]{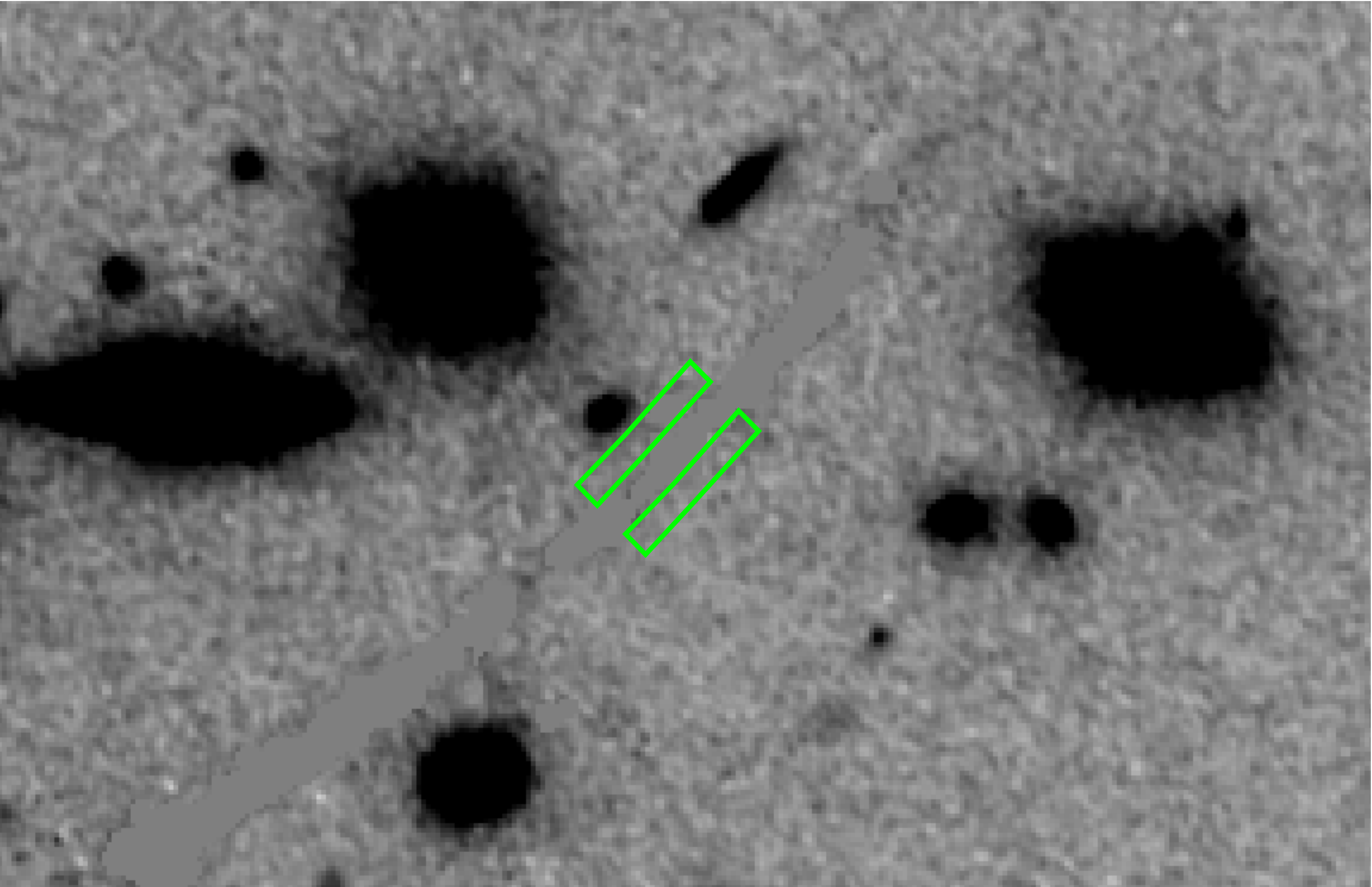}
     \caption{{\bf Diffuse cluster light measurements}
     Stellar surface mass density calculations are performed in the vicinity of the lensed star, within the green boxes shown. The arc and star are masked to avoid contamination, but nearby cluster galaxies are included. This figure shows the {\it HST } F110W band image, which is used to define the extent of the lensed arc.}
     \label{fig:sig_star_region}
 \end{figure}

 The full cluster light and ICL modeling analysis of WHL0137$-$08\ is performed by Jim\'{e}nez-Teja et al. \cite{yoli_whl_icl}.
 Here we mostly care about the stellar mass density in the region around the arc, and more specifically around the lensed star. Thus we focus our measurement on two rectangular regions parallel to the arc, one on each side, extending between the two images of the brightest knot, as shown in Extended Data Figure \ref{fig:sig_star_region}. The arc, knots, and star are all masked from the image prior to measurement, but cluster member galaxies are kept as we are interested in the wings of these galaxies as well as the ICL contribution to stellar density in this region. The extent of the arc is defined in the F110W band, as that is the band in which the arc appears brightest.
 
 The fluxes within these apertures are used to fit a stellar energy distribution (SED). The SED fitting is done using the Fitting and Assessment of Synthetic Templates (FAST) code \cite{Kriek09_fast}. We used stellar population models from Bruzual \& Charlot\cite{Bruzual_Charlot_03}, along with initial mass functions from both Chabrier\cite{ChabrierIMF} and Salpeter\cite{SalpeterIMF} to explore the full range of possible solutions. With this technique, we find a stellar surface mass density of $\Sigma_* \sim 10 ~ M_{\odot} \text{pc}^{-2}$, however the uncertainties allow for values as low as $\sim 1 ~ M_{\odot} \text{pc}^{-2}$ (see Extended Data Table \ref{tab:icl}). Since low stellar surface mass densities introduce more variability in the flux, we explore two regimes with  $\Sigma_* \sim 1 ~ M_{\odot} \text{pc}^{-2}$ and  $\Sigma_* \sim 10 ~ M_{\odot} \text{pc}^{-2}$. Outside this regime, smaller values of  $\Sigma_*$ are unlikely given the observational constraints, and larger values of  $\Sigma_*$ would result in even smaller fluctuations (over time) in the observed flux.

 \subsubsection{Microlensing Simulations} \label{sec:microlens_method}

 To cover the range of possible diffuse light stellar surface mass densities in our microlensing analysis, we ran two simulations of the effects of microlensing on our observed magnification. One simulation assumed a value of $\Sigma_* = 10 ~ M_{\odot} \text{pc}^{-2}$, while the other assumed a value of $\Sigma_* = 1 ~ M_{\odot} \text{pc}^{-2}$. With these two values, we can explore both the high end density estimate, which would produce a denser microlensing caustic network and thus increase the probability that the star would appear at extreme magnification, and the low end estimate which would produce greater variability in the magnification, and a non-negligible probability that one of the two counter-images is unobserved. 
 
 Our simulations follow  Diego et al.\cite{Diego18,Diego19} 
 Since we are assuming the two counter-images form a single unresolved image, in order to compute the total flux, we perform two simulations, one with negative parity and one with positive parity. In both cases, the magnification (in absolute value $|\mu|$) is the same and equal to half the total magnification of the pair of counter-images of the star ($2\mu$). We force the two magnifications to be the same (but opposite sign) by changing slightly the values of $\kappa$ and/or $\gamma$. 
The total flux at a given moment is given by the superposition  of two tracks, one for the simulation with positive parity and one for the simulation with negative parity. Both tracks are forced to have the same orientation with respect to the cluster caustic.  
The very small scale fluctuations observed in the tracks are due to shot noise in the ray-tracing process. 
 
 In the simulation, microlenses are assumed to be point-like, with masses drawn from the mass functions of Spera et al.\cite{Spera15} The mass function is normalized to match our stellar surface mass density measurements around the star. These microlenses are then distributed randomly across a circular region of radius 10 mas, in a lens plane that has a resolution of 20 nas per pixel. 
 For the smooth component, or macromodel, we impose the constraint that the total convergence and shear from the macromodel and the stellar component is consistent with our lens models. The convergence from the smooth component is such that the total magnification is $2\mu \approx 9000$, when the flux from both counter-images is combined into a single unresolved source. In particular, the convergence in the smooth model is determined after fixing the total magnification of each counter-image as $\mu = \mu_\parallel \mu_\perp$,
 where $\mu_\parallel$ and $\mu_\perp$ are the tangential and perpendicular magnifications, respectively.
 This results in a total average magnification (when integrating over long periods of time) of $2\mu=8960$, 
 close to the desired fiducial value $2\mu=9000$.  

\begin{figure}[p!]
     \centering
     \includegraphics[width=0.9\textwidth]{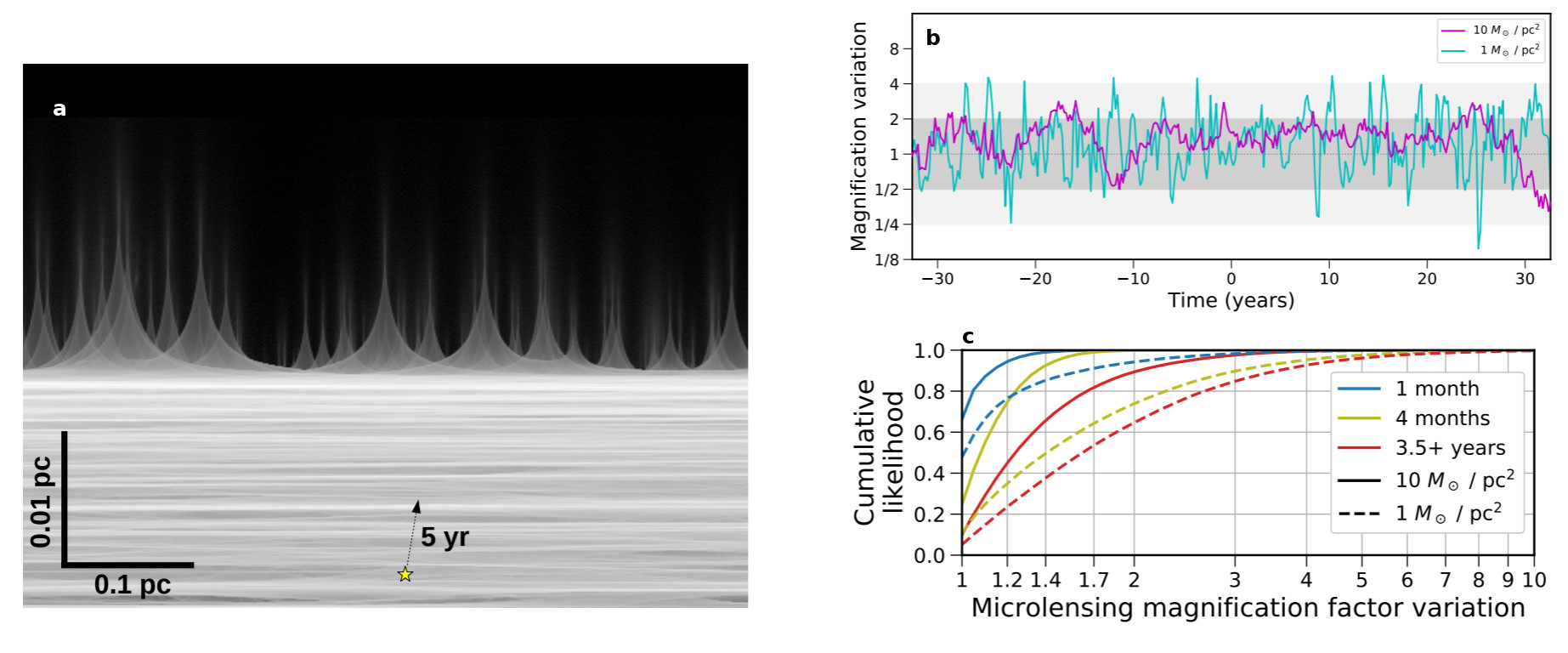}

     \caption{{\bf Flux variations expected from microlensing simulations}
     Microlensing is only expected to vary the total magnification by a factor of 2 -- 3 over time, consistent with the observed steady flux over 3.5 years. Panel {\bf a} shows the simulated microcaustic network arising from stars and stellar remnants within the lensing cluster. The cluster caustic is the extreme magnification horizontal region near the middle of the image, with individual cusps from microlenses still visible beyond the cluster caustic. We estimate Earendel will move relative to the microlens network at $\sim 1000 \text{ km s}^{-1}$ in some unknown direction. Panel {\bf b} shows predicted magnification fluctuations over time arising from this motion in the $1\,M_{\odot} \text{ pc}^{-2}$ case (blue) and the $10\,M_{\odot} \text{ pc}^{-2}$ case (purple), assuming that the relative motion is at an angle of 45\degree. Grey bands highlight a factor of 2 (dark) and a factor of 4 (light) change in magnification. Panel {\bf c} shows the likelihood of magnification variations between two observations separated by different times, again for both the 1 and $10\,M_{\odot} \text{ pc}^{-2}$ cases. Note the \textit{more is less} microlensing effect that reduces variability in the observed images when the density of microlenses increases.}
     \label{fig:microlensing}
 \end{figure}

The magnification in the source plane is then built through a standard ray-tracing method.
The resulting pattern is shown in Extended Data Figure~\ref{fig:microlensing}.
%

 To measure the fluctuations over time, we assume the star is moving at a velocity $v = 1000~\text{km s}^{-1} \sim 0.001$ pc/yr relative to the caustic network\cite{Kelly18,Oguri2018}. 
 This velocity estimate accounts for rotation of the lensed galaxy,
 motions of stars and galaxies within the cluster lens,
 and relative transverse velocities between the cluster lens, lensed galaxy, and Earth with respect to the Hubble flow.
 Windhorst et al.\cite{Windhorst18} test whether $1000~\text{km s}^{-1}$ is reasonable by adding random space motions of up to several thousand $\text{km s}^{-1}$ to well-studied clusters and measuring the effect on the cluster redshift space distributions, finding velocities of $\sim 1000 \text{ km s}^{-1}$ do not distort the observed cluster properties. Thus we infer that this velocity is reasonable in our case. Ultimately, the exact velocity assumed will impact only the duration and frequency of microlensing events, with little to no effect on their amplitude. 
 The direction the star moves relative to the caustic network also impacts the expected variability in magnification. If the star were moving perpendicular to the cluster caustic, we would expect the greatest variation in time, whereas if the star were to move parallel to the cluster caustic we expect much less variation. For our analysis, we assume the star is moving at an angle of 45\degree\ relative to the cluster caustic. This will produce moderate fluctuations in time, with the star typically staying within a factor of two of our measured brightness. As with the velocity, the direction will only impact the duration and frequency of magnification fluctuations. Because we are in a microlensing regime with a larger effective optical depth ($> 1$), microcaustics will overlap and limit the amplitude of variations as the star traverses this caustic network. Thus no matter what velocity and direction we assume, the star will most likely stay within a factor of two of its current magnification, matching our observations.

 Extended Data Figure~\ref{fig:microlensing} shows simulated light curves and likelihoods 
 for both microlensing stellar densities $\Sigma_* = 1$ and 10 $M_{\odot} \text{pc}^{-2}$.
 The higher stellar density reduces the variability in flux as the microlensing caustic network saturates, 
 yielding a consistently high magnification $2\mu \sim 9000$.
 In this case, we can expect with $\sim$65\% confidence that magnifications measured 3.5 years apart will be within a factor of 1.4, as observed.
 If the stellar density is lower (1 $M_{\odot} \text{pc}^{-2}$), this likelihood decreases to $\sim$40\%,
 which is still fairly likely. Therefore, both of these predictions are consistent with our observations.
 
 We also tested a third ``critical'' scenario with maximal time variations and found these very similar to the results for $\Sigma_* = 1 M_{\odot} \text{pc}^{-2}$.
 The degree of variability depends on the product $\mu \Sigma_*$.
 Our simulations had $\mu \Sigma_* = 44800$, 4480, and finally 1600 for the critical case.
 Tighter constraints on \emph{both} parameters $\mu$ and $\Sigma_*$ 
 are required to improve variability predictions.
 Future observations will better constrain these parameters while providing better data on variability or lack thereof. 
 An approved upcoming {\it HST } program (GO 16668; PI Coe) will add time monitoring observations
 to test these predictions and more precisely constrain the baseline flux.

\subsection{Luminosity and Stellar Constraints} \label{sec:stellar}

From our magnification measurements of $\mu = 1000$ -- 40000 derived above, 
and Earendel's observed flux $49 \pm 4$ nJy in the {\it HST } F110W filter (0.9 -- 1.4 $\mu$m),
we calculate a delensed flux of 1 -- 50 pJy, corresponding to an AB magnitude of 38.7 -- 34.7. 
This then gives an absolute UV (1600\AA) magnitude of $-8 < M_{AB} < -12$,
given the distance modulus 48.9 at $z = 6.2$ and flux per unit frequency dimming by $1+z$ (2.1 mag).
We can then calculate the intrinsic stellar luminosity, assuming blackbody spectra for hot stars with effective temperatures $T_{\rm eff} > 40000$ K.
 For cooler stars, 
 we used the lowest surface gravity models available from the grid of empirically-corrected [M/H] $=-1$ 
 stellar atmosphere spectra compiled by Lejeune et al.\cite{Lejeune97}.
 This yields the black tracks in Figure \ref{fig:hr-diagram} for a given magnification and delensed flux.
 Green shaded regions show magnification uncertainties (factor of 7 for each individual lens model);
 the photometric uncertainties are 10\% (insignificant and thus not included).
We also explored the effects of different metallicity stellar atmosphere models (from [M/H] = 0 to -5 using Lejeune et al.\cite{Lejeune97}), and found $\Delta \log L < 0.1$ for $T_{\rm eff} > 10000$ K, and $\Delta \log L \sim 0.2$ for $T_{\rm eff} < 10000$ K, which is insignificant compared to our other uncertainties. 
The redshifted spectrum of a B-type star with temperature $\sim$20000--30000 K maximizes the flux in the F110W filter, whereas spectra of hotter / cooler stars require a higher total bolometric luminosity to produce the same F110W flux.

For our calculations of the stellar luminosity, we assume zero extinction due to interstellar dust. With current data, we cannot robustly estimate the extinction around the star. While BAGPIPES yields an estimate of $A_{V} = 0.15 \pm 0.1$ mag for the full galaxy, we would expect less dust in the ISM near the outskirts of the galaxy where we see the star. On the other hand, we would expect this massive star to still be associated with a star forming region, which would increase the expected dust extinction. To get a rough estimate of the effect of dust, we can assume we have $E(B-V) \sim 0.1$ mag in the region surrounding the star (reasonable for a star cluster in a low-metallicity galaxy\cite{Calzetti15}), and an SMC-like extinction law with $R_V \sim 2.93$. With these assumptions, we would expect a factor $\sim 2$ reduction in flux. This would lead to an equivalent increase in the inferred luminosity. While this is significant, it is still far less than the large magnification uncertainty. 

\begin{figure}[p!]
      \centering
      \includegraphics[width=0.9\textwidth]{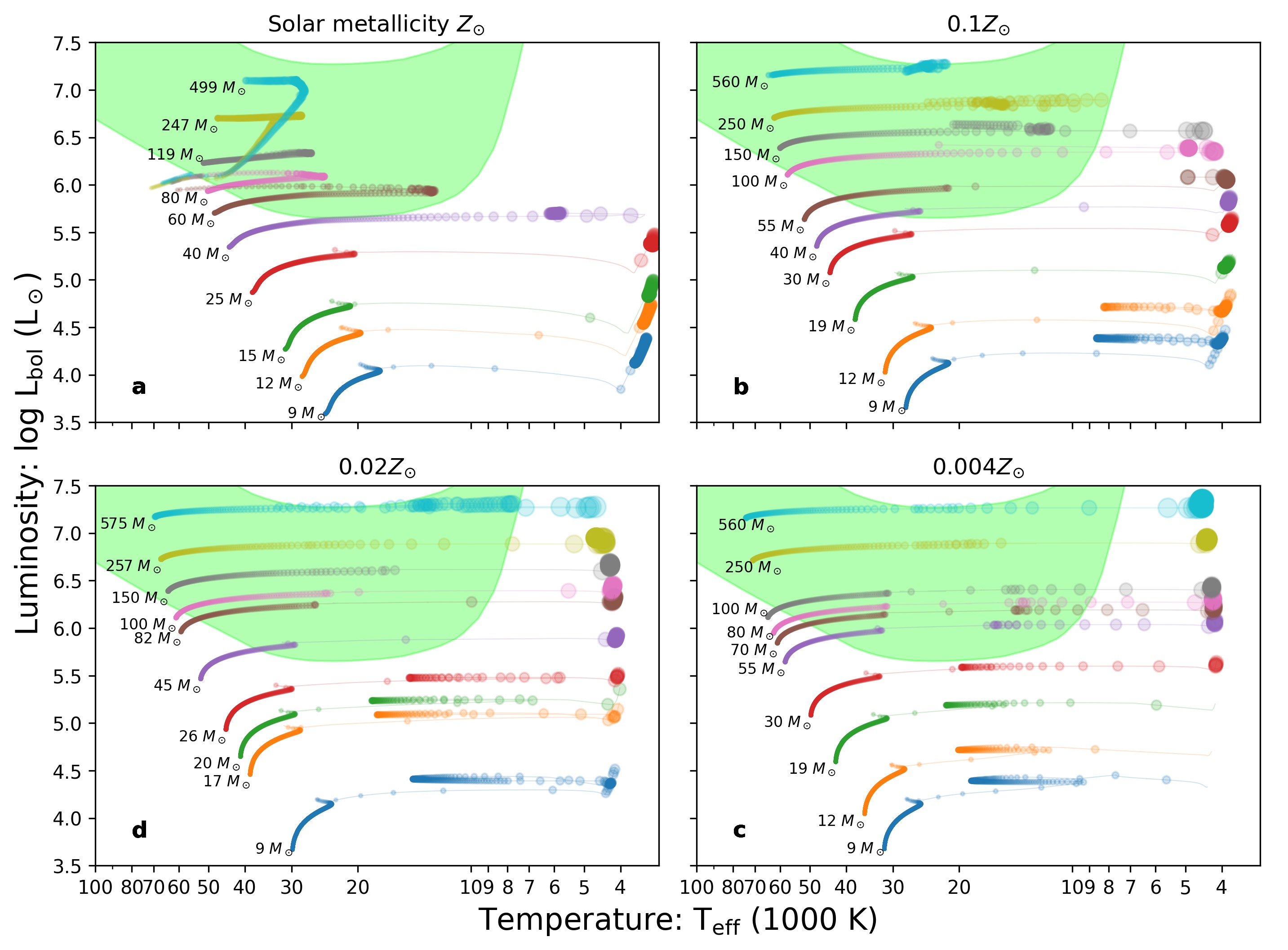}
      \caption{{\bf H-R diagrams with stellar tracks at multiple metallicities}
      A star's metallicity will impact its evolution, so to probe this effect we show here our luminosity constraints compared to stellar tracks from BoOST at metallicities of 1, 0.1, 0.02, and 0.004$Z_{\odot}$ (panels {\bf a, b, c, d} respectively).
      The 0.1$Z_{\odot}$ case is also shown in Figure \ref{fig:hr-diagram},
      and these plots are similar, including the green region allowed by our analysis.
      While the tracks do exhibit some notable differences,
      the resulting mass estimates do not change significantly given the current large uncertainties.}
      \label{fig:4metal}
  \end{figure}

 In Figure \ref{fig:hr-diagram}, we show low-metallicity ($0.1 Z_\odot$) dwarf galaxy predictions from BoOST, as may roughly be expected at $z\sim 6$ based on simulations \cite{Shimizu16}. 
 There is considerable scatter in galaxy metallicities in both observations and simulations. Additionally, the star is not necessarily at the same metallicity as the galaxy overall. To probe the effects of metallicity on our interpretation of the mass of Earendel, we consider a range of stellar tracks from BoOST with varying metallicities. These range from solar metallicity $Z_{\odot}$ down to $0.004 Z_{\odot}$, the full range available from BoOST models. These tracks, along with a green shaded band showing our full luminosity uncertainty across all lens models, is shown in Extended Data Figure \ref{fig:4metal}.
 The differences between various low-metallicity tracks are small relative to our current uncertainties, so the exact choice of metallicity does not impact our conclusions; we still find massive stars best match our constraints. A more important uncertainty is in the stellar modeling of massive stars, which radiate near the Eddington limit. 
 In such stars, the expansion of their outer layers is poorly understood, leading to increased uncertainty on predictions of stellar radii and effective temperatures \cite{Sanyal15}. The luminosity of the star is not impacted by this uncertainty, so these models still provide a useful estimate of the mass of the star.

\begin{figure}[p!]
     \centering
     \includegraphics[width=0.8\textwidth]{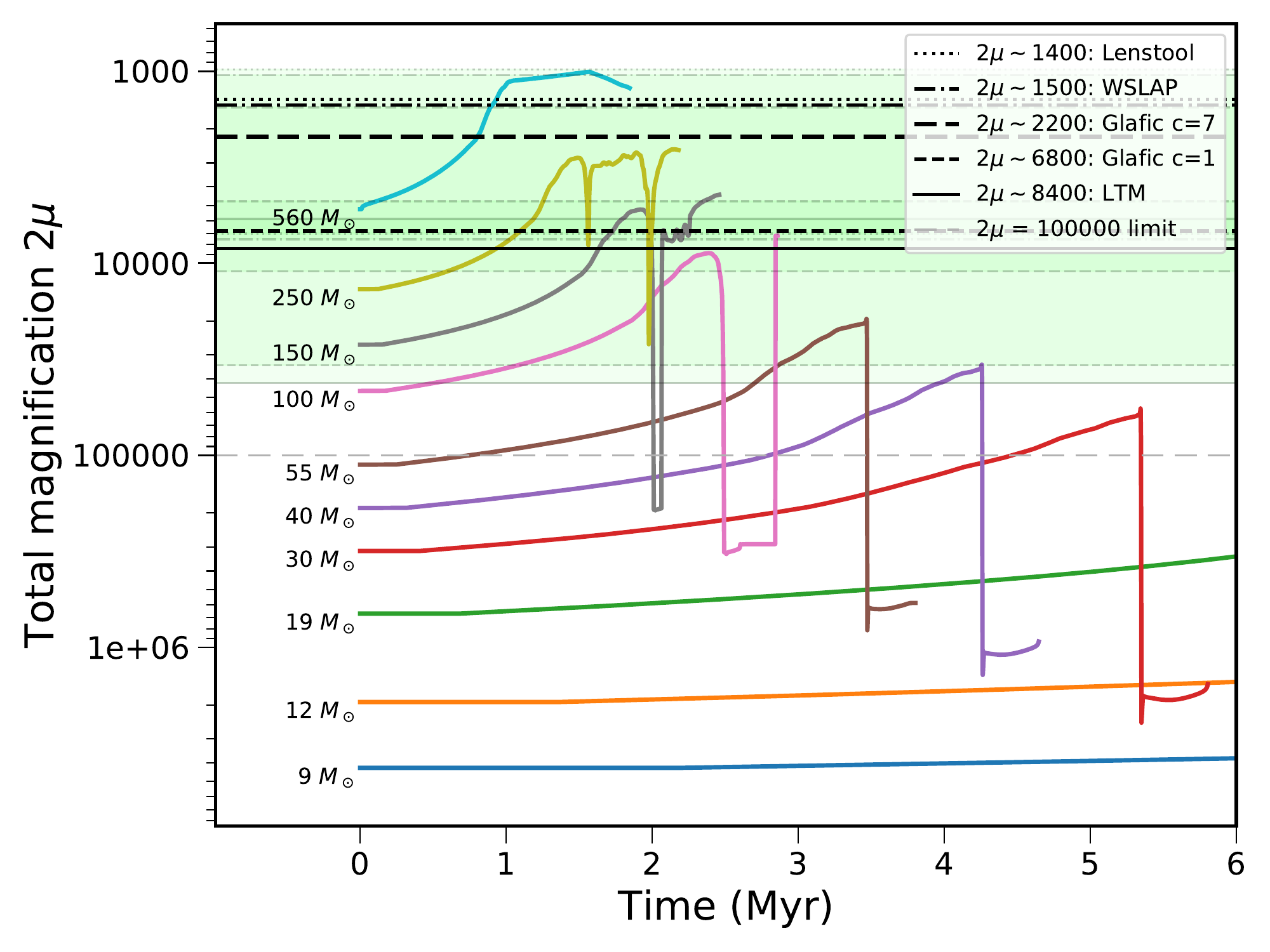}
     \caption{{\bf Stellar evolution tracks versus time}
     Here we show the total magnification required to lens stars to Earendel's apparent magnitude as a function of time on stellar evolution tracks
     (BoOST $0.1 Z_{\odot}$, as plotted in the HR diagram Figure \ref{fig:hr-diagram}).
     This required magnification changes over the lifetime of each star as it varies in luminosity or temperature, changing the flux observed in the F110W filter.
     We find that stars at $\sim 100 M_{\odot}$ and above spend the most time ($\sim$2 Myr) in the green region reproducing Earendel's observed flux, given our magnification estimates.
     But considering that lower mass stars are more numerous, we conclude that masses of roughly $\sim 50 - 100 M_\odot$ are most likely if Earendel is a single star.
     }
     \label{fig:star_time}
 \end{figure}

 In Extended Data Figure \ref{fig:star_time}, we show the BoOST $0.1 Z_{\odot}$ stellar evolution tracks vs. time.
 We see that very massive stars of $100 M_{\odot}$ or more spend the greatest time ($\sim 2$ Myr) with a luminosity matching Earendel within the uncertainties. The next less massive track $\sim 55 M_{\odot}$ would only match our luminosity constraint for $\sim 0.5$ Myr. This shorter time would decrease the probability of observing such a star by $\sim 1/4$. On the other hand, lower mass $55 M_{\odot}$ stars may be roughly 4 times as numerous as $100 M_{\odot}$ stars, depending on the IMF. From this simple analysis, we estimate that Earendel's light may be most likely generated by a star with $\sim 50 - 100 M_\odot$. More massive stars are less likely because they are less numerous, while a less massive star would not be bright enough. Given the large uncertainties on our observational constraints, we leave more detailed analysis of lifetimes, formation rates, and magnification probabilities for future work.

 Such massive stars are rarely single \cite{Sana12,Sana14}.
 Multiple less massive stars could also combine to produce the observed luminosity.
 While it is most likely that a single star will dominate the light in such a system, there is a possibility of finding tightly bound multiple systems of similar masses, and thus similar brightnesses. 
 El-Badry et al.\cite{ElBadry19_twinsies} find a sharp excess of ``twin" systems, with a mass ratio $\gtrsim 0.95$ indicating that the stars are roughly equal masses. 
 This analysis was restricted to lower mass stars, but Moe \& Di Stefano\cite{Moe17} find a similar (although more modest) excess for more massive systems. 
 Additionally, Moe \& Di Stefano\cite{Moe17} find that the fraction of stars in triple and quadruple systems increases as the primary star's mass increases, up to a quadruple fraction of $\sim 50$\% for stars of mass $\sim 25 M_{\odot}$. 
 In any case of a twin/triple/quadruple star,
 the contribution from companion stars becomes non-negligible. 
 This would effectively reduce the inferred mass of the primary star, potentially down to $\sim 20 M_{\odot}$ in the case of a quadruple system of equal mass main sequence stars. 
 We note that multiple bright stars would also dampen microlensing variations,
 as one star may be crossing a microcaustic while others are not,
 further supporting our observation of relatively stable flux.

  With the {\it HST } photometry available, we cannot reliably distinguish between different stellar types and effective temperatures, and thus stellar mass. We have multi-band imaging, but only the F110W band has a reliable ($> 5\sigma$ significance) detection. Other WFC3-IR bands do detect the star, but at much lower significance, the highest being F160W with a $\sim 3 \sigma$ detection. Additionally, these other IR bands were taken in the original RELICS imaging. The 3.5 year gap between observations provides ample time for the magnification to vary considerably. Our microlensing analysis suggests that the magnification will stay high ($\mu \geq 1000$) for many years, but fluctuations of a factor of 2 are expected. 
 
 ACS/F814W imaging was obtained in every epoch of observations, however the stacked detection in this band is still only at $4.4\sigma$ confidence. Additionally, this bandpass spans the Lyman-$\alpha$ break at $z \sim 6$, so the F814W flux is primarily a function of redshift: more flux drops out as redshift increases up to $z=7$.
 Any SED constraint from {\it HST } photometry would be weak. 
 Future spectroscopic observations with our approved JWST program (GO 2282; PI Coe) will determine the type and temperature of this star, placing it on the H-R diagram.

 \subsection{Probability of Observing a Massive Star} \label{sec:probability} 
 
 In the following, we assess the probability of observing a lensed star with $M_* \gtrsim 100 M_{\odot}$ at sufficiently high magnification to be detected in our data. To do so, we use our model of the Sunrise Arc to estimate the total star formation rate (SFR) of the host galaxy within a region close to the lensing caustic, and use assumptions on the stellar IMF to convert this into an estimate on the number of high-mass stars within this area. 
Using the formalism presented by Diego\cite{Diego19}, and fitting for necessary constants using our LTM lens model, we calculate a source plane area of $\sim 500 \text{ pc}^2$ ($\sim 0.5$ pc perpendicular to the caustic times $\sim 1000$ pc along the caustic) that intersects the host galaxy at a magnification of $2\mu\gtrsim 4200$ (set by the minimum magnification to detect a $M_* \sim 100 \ M_\odot$ star in our data). We next estimate the surface brightness of the arc close to the location of the star, finding $I \sim 6\times 10^{-6} \text{ nJy pc}^{-2}$. By assuming that the surface brightness of the host galaxy remains approximately constant along the caustic, we may then convert surface brightness to total SFR within the $2\mu > 4200$ region that intersects the host galaxy. Using Starburst99 v.7.0.1 \cite{Leitherer99} under the assumption of a Kroupa\cite{Kroupa01_imf} IMF throughout the 0.01--120 $M_\odot$ interval and a constant-SFR stellar population with $Z=0.001$ at age 100 Myr, we find a total $SFR \sim 2\times 10^{-4} ~ M_{\odot}\text{ yr}^{-1}$ in the host galaxy region magnified by $2\mu > 4200$.

We next calculate the probability of observing a star of mass $\geq 100\ M_\odot$ at this SFR. This is done with two star formation prescriptions, one assuming clustered star formation wherein stars are distributed into star clusters with a cluster mass function $\mathrm{d}N_\mathrm{cluster}/\mathrm{d}M_\mathrm{cluster} \propto M_\mathrm{cluster}^{-2}$ between 20 -- $10^7\ M_\odot$, then randomly sampled from the IMF with the limit $M_* < M_\mathrm{cluster}$ using the \text{SLUG} v2.0 code \cite{daSilva12,Krumholz15}. This tends to result in fewer high mass stars. The other possibility is unclustered star formation, in which stars are randomly sampled from the IMF without first being split into clusters. This method results in a greater proportion of massive stars. From this calculation, we find a probability $P(\geq 100 M_{\odot}) \sim 2$\% in the clustered scenario, and $\sim 4$\% in the unclustered scenario.

From this calculation, we conclude that one might expect to find such a massive, lensed star in about 1 in 25--50 such caustic crossing galaxies.
Tens of galaxies like this have been observed in {\it HST } images from various programs.
Therefore, the probability of such a discovery is reasonable.

We note that a different choice of IMF over the stated mass range could impact the calculated SFR, up to a factor $\sim 1.5$ if we were to use the Salpeter\cite{SalpeterIMF} IMF \cite{MadauDickinson14}. However, this change will be largely canceled out by the lower probability per unit stellar mass of forming $> 100 M_{\odot}$ stars using the Salpeter IMF. The difference if we were to use a Chabrier\cite{ChabrierIMF} IMF would be even smaller, as that is more similar to the Kroupa IMF over our mass range. Ultimately, any uncertainty introduced by the choice of IMF will be sub-dominant compared to other assumptions, such as the assumed metallicity. 
However, if the stellar IMF would be more top-heavy than the Kroupa IMF (i.e., contain a larger fraction of massive stars), as has been argued to be the case for star formation in low-metallicity environments \cite{Kehrig18}, then the probability for detecting a $\geq 100 \ M_\odot$ star could be significantly higher. The probability would also increase in scenarios where the host galaxy, despite being metal-enriched, contains a fraction of Pop III stars, as in the simulations by Sarmento et al.\cite{Sarmento18,Sarmento19}. 
From the MESA mass-luminosity relation of such stars, Windhorst et al.\cite{Windhorst18} showed that most of their light comes from $20-200 \ M_\odot$ stars, so that finding a single $\sim 100 \ M_\odot$ star during a significant caustic magnification is possible.

 \subsection{Alternative Possibilities}
 
 There are a few alternative possibilities we consider for this object. One such possibility is that Earendel is a Population III star with zero metallicity. Calculations of observable properties of Pop III stars from \texttt{MESA} stellar evolution models show that a $> 50 M_{\odot}$ in a hydrogen-depleted phase would match our delensed flux constraint, as would a ZAMS star of $> 300 M_{\odot}$ \cite{Windhorst18}.
 
 
 The lifetime of a massive Pop III star would be short relative to its host galaxy.
 It would therefore require a pristine zero-metallicity environment within the host galaxy to form.
 Such regions become less common as more early generation stars explode and enrich their surroundings with heavier elements. 
 Our SED fitting of the Sunrise Arc gives a stellar mass of $M_* \sim 3\times 10^7 M_{\odot}$, 
 from which we might expect a metallicity on the order of $0.01Z_{\odot}$ or $0.1Z_{\odot}$ \cite{Shimizu16}.
 This non-zero metallicity would indicate that some enrichment has taken place.
 Even if we assume the galaxy overall has been enriched, finding a Pop III stellar population is not ruled out. 
 Some models predict that pockets of zero-metallicity gas (from which Pop III stars can form) may still exist at $z \sim 6$, particularly near the outskirts of galaxies \cite{Trenti09}. 
 Observationally, Vanzella et al.\cite{Vanzella2020} report a strongly lensed star cluster consistent with being a complex of Pop III stars at $z = 6.6$. 
 In this case, pockets of zero-metallicity stars may exist in otherwise metal enriched galaxies. 
 Spectroscopic follow-up will be required to assess the possibility of Earendel being a Pop III star. 
 
 Although the probability of Earendel being a zero-metallicity Pop III star is low, if it turns out to be such an object it would be the first such star observed. 
 This would provide important confirmation that such stars formed, and would offer an incredible opportunity to study one in detail. 
 Furthermore, such a star is a possible progenitor for the recently observed binary black hole merger GW190521, which is too massive to be explained by standard stellar remnants \cite{ligo_190521}. 
 Recent studies \cite{farrell20_lowZbbh, kinugawa20_pop3bbh} have proposed that extremely metal poor or zero metallicity stars are viable progenitors for this event. 
 Finding such a star would offer a chance to study it in detail and refine models of how these stars collapse into black holes.

 Another possibility is that this object is an accreting stellar mass black hole. 
 If the black hole were persistently fed by a lower mass star overfilling its Roche lobe, it could continue to shine up to 60 Myr \cite{Windhorst18}. Note we assume this would be a persistent source. A transient outburst, which would shine for weeks to months \cite{zdziarski-gierlinski04_bhxb}, would be ruled out by the lack of variation observed (see \S\ref{sec:variability}). 
 A stellar mass black hole accretion disk would have strong X-ray emission, whereas a star would not. 
 Following the multicolor accretion disk model in Windhorst et al.\cite{Windhorst18}, a $200 M_{\odot}$ black hole formed from a $\sim 300 M_{\odot}$ Pop III star would have an inner accretion disk with $T_{max} \sim$ 7.7 keV $\sim 3\times 10^7$ K. 
 Most of the X-ray emission would originate near the center of the disk, from radii around a few times the Schwarzschild radius ($\gtrsim$ 900--1800 km). 
 The maximum possible magnification of the tiny X-ray emitting region could then be substantially (perhaps 100$\times$) larger than that for the rest-frame UV stellar light, or $\mu \gtrsim 10^6$. 
 Most of the remainder of the accretion disk would shine in the rest-frame UV at the Eddington limit with very similar size, luminosity, and surface brightness as a massive star. 
 Hence, a stellar mass black hole accretion disk would appear very similar in the HST images to a massive star.
 Analysis of archival X-ray data from XMM-Newton showed no clear signal near this position, supporting the stellar interpretation. 
 We note, however, that the 6\arcsec\ spatial resolution of XMM-Newton would dilute the signal from such a black hole accretion disk. 
 Deeper, higher-resolution X-ray images with the Chandra X-ray Observatory (resolution $\sim 0.5''$) or the upcoming Athena mission could conclusively determine if this source is a black hole.

We also consider the possibility that this object is not associated with the Sunrise Arc, and thus not a lensed star at $z \sim 6$. 
The first possibility would be a local star which happens to align with the background arc.
While possible, this is unlikely given the exact alignment of the object with the background arc.
Holwerda et al.\cite{Holwerda14} found $\sim 1.2$ M-type dwarfs per arcmin$^2$ out to magnitude 24 in their analysis of multiple fields observed with HST.
Rescaling to 27th magnitude, we estimate we might observe of order $\sim 100$ such stars per arcmin$^2$ in our observations.
Given the small solid angle surrounding the critical curve crossings in the Sunrise Arc (constrained to within 0.1\arcsec), the probability of one of these local dwarfs aligning with both the arc and critical curve by chance is of order 0.01\%. 
If Earendel were a local brown dwarf, we might see some evidence of proper motion over the 3.5 year observation window. 
We see no evidence of motion, strengthening the interpretation that this star is associated with the Sunrise Arc. 
We note that we cannot conclusively rule out a brown dwarf based on existing {\it HST } photometry. We fit the SED of Earendel alone to brown dwarf spectra from the SpeX Prism Library \cite{spex_browndwarf}, and find that a 3000 K local star could reproduce our observations. 
We expect upcoming {\it JWST }\ photometry and spectroscopy to rule out a brown dwarf conclusively.
For now, we rely on the unlikely chance alignment and lack of proper motion to disfavor a local brown dwarf. 

Other possibilities to consider for this object are distinct galaxies in the foreground, cluster, or lensed background.
But even analyzed independently, Earendel's photometric redshift is the same as the full galaxy: $z = 6.2 \pm 0.1$ (95\% CL) with negligible likelihood at lower redshifts according to BPZ given the Lyman break, which is clear even in this faint object.
Furthermore, any foreground / lensed background galaxy would most likely appear larger and spatially resolved in the {\it HST }\ images.
Note a $z < 6$ background galaxy, say at $z \sim 2$, would not be on the lensing critical curve for that redshift, but it would still be magnified by a factor of a few to perhaps tens, requiring a small galaxy to not appear spatially resolved when magnified.
A quasar would still appear point-like when lensed, but of course quasars are less numerous and we would expect a redder rest-frame ultraviolet continuum slope \cite{Hainline11_agn}. 
Again, we expect upcoming to {\it JWST } observations to spectroscopically confirm Earendel is a star at $z \sim 6.2$ within the Sunrise Arc galaxy.

\section{Data Availability}

All {\it HST } image data used in this analysis are publicly available on the Mikulski Archive for Space Telescopes (MAST), and can be found through the following DOI hyperlinks:
\href{https://doi.org/10.17909/T9SP45}{RELICS}, and follow-up
\href{https://doi.org/10.17909/t9-ztav-b843}{HST GO 15842}.



 \section{Acknowledgements}

The RELICS Hubble Treasury Program (GO 14096) and follow-up program (GO 15842) 
consist of observations obtained by the NASA/ESA \emph{Hubble Space Telescope (HST)}.
Data from these {\it HST } programs were obtained from the Mikulski Archive for Space Telescopes (MAST), operated by the Space Telescope Science Institute (STScI).
Both {\it HST } and STScI are operated by the Association of Universities for Research in Astronomy, Inc.~(AURA), under NASA contract NAS 5-26555.
The {\it HST } Advanced Camera for Surveys (ACS) was developed under NASA contract NAS 5-32864.

JMD acknowledges the support of project PGC2018-101814-B-100 (MCIU/AEI/MINECO/FEDER, UE)  and Mar\'ia de Maeztu, ref. MDM-2017-0765. 

AZ acknowledges support from the Ministry of Science \& Technology, Israel. 

RAW acknowledges support from NASA JWST Interdisciplinary Scientist
grants NAG5-12460, NNX14AN10G and 80NSSC18K0200 from GSFC.

EZ and AV acknowledge funding from the Swedish National Space Board.

MO acknowledges support from World Premier International Research Center Initiative, MEXT, Japan, and JSPS KAKENHI Grant Number JP20H00181, JP20H05856, JP18K03693.

GM received funding from the European Union’s Horizon 2020 research and innovation programme under the Marie Skłodowska-Curie grant agreement No MARACAS - DLV-896778.

PLK acknowledges support from NSF AST-1908823.

Y.~J-T acknowledges financial support from the European Union’s Horizon 2020 research and innovation programme under the Marie Skłodowska-Curie grant agreement No 898633, and from the State Agency for Research of the Spanish MCIU through the
"Center of Excellence Severo Ochoa" award to the Instituto de Astrof\'isica de Andaluc\'ia (SEV-2017-0709).

S.T. acknowledges The Cosmic Dawn Center of Excellence is funded by the Danish National Research Foundation under grant No. 140.

\section{Author Information}

\subsection{Contributions}
B.W. identified the star, led the lens modeling and size constraint analysis, and wrote the majority of the manuscript.
D.C. proposed and carried out observations, measured photometry and redshifts, and helped analyze size and magnification constraints.
J.M.D. performed and analyzed microlensing simulations, and contributed to lens model analysis.
A.Z., G.M., M.O., and K.S. contributed to the lens model analyses.
E.Z. calculated stellar constraints based on observed magnitude and magnification. 
P.D. and Y.J.-T. calculated stellar surface mass densities used in microlensing analysis. 
P.K. and R.W. helped compare results and methods to previous lensed star detections and theoretical predictions
F.X.T., S.E.d.M., and A.V. contributed to stellar constraint analysis and interpretation.
R.J.A. reduced the HST images.
M.B. and V.S. obtained and analyzed Spitzer data.
All authors contributed to the scientific interpretation of the results and to aspects of the analysis and writing.

\subsection{Corresponding author}
Correspondence to Brian Welch: \href{mailto:bwelch7@jhu.edu}{bwelch7@jhu.edu}.

\section{Ethics Declarations}
{\bf Competing Interests}

The authors declare no competing interests. 

\begin{table}[p!]
     \centering
     \begin{tabular}{c c c c c c c}
         &  & HST GO 14096 & HST GO 15842 & Depth 5$\sigma$ & Sunrise Arc & Earendel \\
        Camera & Filter & Exposure Time (s) & Exposure Time (s) & (AB mag) & Flux (nJy) & Flux (nJy)\\
        \hline 
        ACS     & F435W & 2072 &       & 27.2 & $-69 \pm 56$ & $-8 \pm 12$ \\
        ACS     & F475W &      &  3988 & 27.9 & $ 16 \pm 27$ & $9 \pm 6$  \\
        ACS     & F606W & 2072 &       & 27.6 & $-51 \pm 33$ & $-2 \pm 7$  \\
        ACS     & F814W & 2243 & 11083 & 28.0 & $312 \pm 21$ & $19 \pm 4$  \\
        WFC3/IR & F105W & 1411 &       & 26.7 & $1321 \pm 74$ & $30 \pm 14$  \\
        WFC3/IR & F110W &      &  5123 & 27.7 & $1187 \pm 21$ & $49 \pm 4$  \\
        WFC3/IR & F125W &  711 &       & 26.0 & $1351 \pm 137$ & $37 \pm 26$  \\
        WFC3/IR & F140W &  711 &       & 26.2 & $1197 \pm 109$ & $21 \pm 21$  \\
        WFC3/IR & F160W & 1961 &       & 26.5 & $1088 \pm 74$ & $46 \pm 15$  \\
     \end{tabular}
     \caption{Hubble imaging of WHL0137-08 in nine filters 
     and photometry measured for the Sunrise Arc and Earendel.
     Imaging was obtained by RELICS (HST GO 14096) and follow-up imaging 3.5 years later (HST GO 15842).
     Final 5$\sigma$ depths for point sources are given in column 5.
     Fluxes used in SED fitting and plotted in Figures \ref{fig:photometry} \& \ref{fig:variation} are given in column 6 (full arc) and column 7 (Earendel individually), along with 68\% confidence uncertainties.}
     \label{tab:HST}
 \end{table}
 
  \begin{table}[p!]
     \centering
     \begin{tabular}{c|c}
          & Stellar Mass Density \\
        IMF  & ($M_{\odot} ~ \text{pc}^{-2}$) \\
        \hline 
         Chabrier & 8 $\! \left[ _{0.7} ^{17} \right] \!$ \\
         Salpeter & 15 $\! \left[ _{1.0} ^{30} \right] \!$
     \end{tabular}
     \caption{Stellar surface mass densities from two possible IMFs. These values include both ICL and the wings of cluster member galaxies. Most likely values are followed by 68\% confidence ranges in brackets.}
     \label{tab:icl}
 \end{table}

\begin{thebibliography}{99}

\bibitem[1]{RiveraThorsen17_sunburst} Rivera-Thorsen, T.E. \emph{et al.} The Sunburst Arc: Direct Lyman $\alpha$ escape observed in the brightest known lensed galaxy. \emph{Astron. Astrophys.} \textbf{608,} L4 (2017).

\bibitem[2]{Johnson17L} Johnson, T. L. \emph{et al.} Star Formation at z = 2.481 in the Lensed Galaxy SDSS J1110+6459: Star Formation Down to 30 pm Scales. \emph{Astrophys. J. Lett.} {\bf 843,} L21 (2017).

\bibitem[3]{Kelly18} Kelly, P. L. \emph{et al.} Extreme magnification of an individual star at redshift 1.5 by a galaxy-cluster lens. \emph{Nature Astronomy} {\bf 2,} 334-342 (2018).

\bibitem[4]{Rodney18} Rodney, S. A. \emph{et al.} Two peculiar fast transients in a strongly lensed host galaxy. \emph{Nature Astronomy} {\bf 2,} 324-333 (2018).

\bibitem[5]{Chen19} Chen, W. \emph{et al.} Searching for Highly Magnified Stars at Cosmological Distances: Doscovery of a REdshift 0.94 Supergiant in Archival Images of the Galaxy Cluster MACS J0416.1-2403. \emph{Astrophys. J.} {\bf 881,} 8 (2019).

\bibitem[6]{Kaurov19_lensedstar} Kaurov, A. A., Dai, L., Venumadhav, T., Miralda-Escud\'e, J. \& Frye, B. Highly Magnified Stars in Lensing Clusters: New Evidence in a Galaxy Lensed by MACS J0416.1-2403. \emph{Astrophys. J.} {\bf 881,} 58 (2019).

\bibitem[7]{Coe19_relics} Coe, D. \emph{et al.} RELICS: Reionization Lensing Cluster Survey. \emph{Astrophys. J.} {\bf 884,} 85 (2019).

\bibitem[8]{Salmon2020} Salmon, B. \emph{et al.} RELICS: The Reionization Lensing Cluster Survey and the Brightest High-z Galaxies. \emph{Astrophys. J.} {\bf 889,} 189 (2020).

\bibitem[9]{RivThor19_sunburst2} Rivera-Thorsen, T. E. \emph{et al.} Gravitational lensing reveals ionizing ultraviolet photons escaping from a distant galaxy. \emph{Science} {\bf 366,} 738-741 (2019). 

\bibitem[10]{Zitrin15} Zitrin, A. \emph{et al.} Hubble Space Telescope Combined Strong and Weak Lensing Analysis of the CLASH Sample: Mass and Magnification Models and Systematic Uncertainties. \emph{Astrophys. J.} {\bf 801,} 44 (2015).

\bibitem[11]{Zitrin09} Zitrin, A. \emph{et al.} New multiply-lensed galaxies identified in ACS/NIC3 observations of Cl0024+1654 using an improved mass model. \emph{Mon. Not. R. Astron. Soc.} {\bf 395,} 1319-1332 (2009).

\bibitem[12]{Broadhurst05} Broadhurst, T. \emph{et al.} Strong-Lensing Analysis of A1689 from Deep Advanced Camera Images. \emph{Astrophys. J.} {\bf 621,} 53-88 (2005).

\bibitem[13]{JulloLenstool09} Jullo, E., \& Kneib, J. P. Multiscale cluster lens mass mapping - I. Strong lensing modelling. \emph{Mon. Not. R. Astron. Soc.} {\bf 395,} 1319-1332 (2009).

\bibitem[14]{JulloLenstool07} Jullo, E., \emph{et al.} A Bayesian approach to strong lensing modelling of galaxy clusters. \emph{New Journal of Physics} {\bf 9,} 447 (2007).

\bibitem[15]{Oguri2010} Oguri, M. The Mass Distribution of SDSS J1004+4112 Revisited. \emph{PASJ} {\bf 62,} 1017 (2010).

\bibitem[16]{Diego07wslap2} Diego, J. M., Tegmark, M., Protopapas, P. \& Sandvik, H. B. Combined reconstruction of weak and strong lensing data with WSLAP. \emph{Mon. Mot. R. Astron. Soc.} {\bf 375,} 958-970 (2007).

\bibitem[17]{Diego05wslap} Diego, J. M., Protopapas, P., Sandvik, H. B. \& Tegmark, M. Non-parametric inversion of strong lensing systems. \emph{Mon. Not. R. Astron. Soc.} {\bf 360,} 477-491 (2005).

\bibitem[18]{Diego19} Diego, J. M. The Universe at extreme magnification. \emph{Astron. Astrophys.} {\bf 625,} A84 (2019).

\bibitem[19]{Meneghetti17} Meneghetti, M. \emph{et al.} The Frontier Fields lens modelling comparison project. \emph{Mon. Mot. R. Astron. Soc.} {\bf 472,} 3177-3216 (2017).

\bibitem[20]{Venumadhav2017} Venumadhav, T., Dai, L. \& Miralda-Escud\'e, J. Microlensing of Extremely Magnified Stars near Caustics of Galaxy Clusters. \emph{Astrophys. J.} {\bf 850,} 49 (2017).

\bibitem[21]{Diego18} Diego, J. M. \emph{et al.} Dark Matter under the Microscope: Constraining Compact Dark Matter with Caustic Crossing Events. \emph{Astrophys. J.} {\bf 857,} 25 (2018). 

\bibitem[22]{Dai2021} Dai, L. Statistical microlensing towards magnified high-redshift star clusters. \emph{Mon. Mot. R. Astron. Soc.} {\bf 501,} 5538-5553 (2021). 

\bibitem[23]{Portegies-zwart2010_YMCs} Portegies Zwart, S. F., McMillan, S. L. W. \& Gieles, M. Young Massice Star Clusters. \emph{Annu. Rev. Astron. Astrophys.} {\bf 48,} 431-493 (2010).

\bibitem[24]{arches_figer99} Figer, D. F., McLean, I. S. \& Morris, M. Massive Stars in the Quintuplet Cluster. \emph{Astrophys. J.} {\bf 514,} 202-220 (1999).

\bibitem[25]{Bouwens17} Bouwens, R. J. \emph{et al.} Very low-luminosity galaxies in the early universe have observed sizes similar to single star cluster complexes. \emph{arXiv e-prints,} arXiv:1711.02090 (2017).

\bibitem[26]{Vanzella19_13pc} Vanzella, E. \emph{et al.} Massive star cluster formation under the microscope at z = 6. \emph{Mon. Not. R. Astron. Soc.} {\bf 483,} 3618-3635 (2019).

\bibitem[27]{Behrendt19_clumps} Behrendt, M., Schartmann, M. \& Burkert, A. The possible hierarchical scales of observed clumps in high-redshift disc galaxies. \emph{Mon. Not. R. Astron. Soc.} {\bf 488,} 306-323 (2019).

\bibitem[28]{Sana12} Sana, H. \emph{et al.} Binary Interaction Dominates the Evolution of Massive Stars. \emph{Science} {\bf 337,} 444 (2012).

\bibitem[29]{Sana14} Sana, H. \emph{et al.} Southern Massive Stars at high Angular Resolution: Observational Campaign and Companion Detection. \emph{Astrophys. J. Suppl.} {\bf 215,} 15 (2014).

\bibitem[30]{Moe17} Moe, M. \& Di Stefano, R. Mind Your Ps and Qs: The Interrelation between Period (P) and Mass-ratio (Q) Distributions of Binary Stars. \emph{Astrophys. J. Suppl.} {\bf 230,} 15 (2017).

\bibitem[31]{Szecsi20_boost} Sz\'ecsi, D., W\"unsch, R., Agrawal, P. \& Langer, N. `Bonn' Optimized Stellar Tracks (BoOST). Simulated Populations of Massive and Very Massive Stars for Astrophysical Applications. \emph{arXiv e-prints,} arXiv:2004.08203 (2020).

\bibitem[32]{Shimizu16} Shimizu, I., Inoue, A. K., Okamoto, T. \& Yoshida, N. Nebular line emission from z $>$ 7 galaxies in a cosmological simulation: rest-frame UV to optical lines. \emph{Mon. Not. R. Astron. Soc.} {\bf 461,} 3563-3575 (2016).

\end{thebibliography}

\begin{thebibliography}{99}

\bibitem[33]{WHL12} Wen, Z. L., Han, J. L. \& Liu, F. S. A Catalog of 132684 Clusters of Galaxies Identified from Sloan Digital Sky Survey III. \emph{Astrophys. J. Suppl.} {\bf 199,} 34 (2012).

\bibitem[34]{WenHan15} Wen, Z. L. \& Han, J. L. Calibration of the Optical Mass Proxy for Clusters of Galaxies and an Update of the WHL12 Cluster Catalog. \emph{Astrophys. J.} {\bf 807,} 178 (2015).

\bibitem[35]{SDSS3} Alam, S. \emph{et al.} The Eleventh and Twelfth Data Releases of the Sloan Digital Sky Survey: Final Data from SDSS-III. \emph{Astropys. J. Suppl.} {\bf 219,} 12 (2015).

\bibitem[36]{PSZ2} Planck Collaboration \emph{et al.} Planck 2015 results. XXVII. The second Planck catalogue of Sunyaev-Zeldovich sources. \emph{Astron. Astrophys.} {\bf 594,} A27 (2016).

\bibitem[37]{SZ} Sunyaev, R. A., \& Zeldovich, Y. B. Small-Scale Fluctuations of Relic Radiation. \emph{Ap\&SS} {\bf 7,} 3-19 (1970).

\bibitem[38]{Strait21} Strait, V. \emph{et al.} RELICS: Properties of z $\geq$ 5.5 Galaxies Inferred from Spitzer and Hubble Imaging, Including A Candidate z $\sim$ 6.8 Strong [OIII] emitter. \emph{Astrophys. J.} {\bf 910,} 135 (2021).

\bibitem[39]{sextractor} Bertin, E. \& Arnouts, S. SExtractor: Software for source extraction. \emph{Astron. Astrophys. Suppl.} {\bf 117,} 393-404 (1996).

\bibitem[40]{Benitez00bpz} Beintez, N. Bayesian Photometric Redshift Estimation. \emph{Astrophys. J.} {\bf 536,} 571-583 (2000).

\bibitem[41]{Coe06bpz} Coe, D. \emph{et al.} Galaxies in the Hubble Ultra Deep Field. I. Detection, Multiband Photometry, Photometric Redshifts, and Morphology. \emph{Astron. J.} {\bf 132,} 926-959 (2006).

\bibitem[42]{Carnall18_bagpipes} Carnall, A. C., McLure, R. J., Dunlop, J. S. \& Dav\'e, R. Inferring the star formation histories of massive quiescent galaxies with BAGPIPES: evidence for multiple quenching mechanisms. \emph{Mon. Not. R. Astron. Soc.} {\bf 480,} 4379-4401 (2018). 

\bibitem[43]{Eldridge17_bpass} Eldridge, J. J. \emph{et al.} Binary Population and Spectral Synthesis Version 2.1: Construction, Observational Verification, and New Results. \emph{PASA} {\bf 34,} e058 (2017).

\bibitem[44]{Ferland17} Ferland, G. J. \emph{et al.} The 2017 Release Cloudy. \emph{RMxAA} {\bf 53,} 385-438 (2017).

\bibitem[45]{SalpeterIMF} Salpeter, E. E. The Luminosity Function and Stellar Evolution. \emph{Astrophys. J.} {\bf 121,} 161 (1955).

\bibitem[46]{Calzetti00} Calzetti, D. \emph{et al.} The Dust Content and Opacity of Actively Star-forming Galaxies. \emph{Astrophys. J.} {\bf 533,} 682-695 (2000).

\bibitem[47]{Ellis97} Ellis, R. S. \emph{et al.} The Homogeneity of Spheroidal Populations in Distant Clusters. \emph{Astrophys. J.} {\bf 483,} 582-596 (1997).

\bibitem[48]{Stanford98} Stanford, S. A., Eisenhardt, P. R. \& Dickinson, M. The Evolution of Early-Type Galaxies in Distant Clusters. \emph{Astrophys. J.} {\bf 492,} 461-479 (1998).

\bibitem[49]{Hastings1970} Hastings, W. K. Monte Carlo sampling methods using Markov chains and their applications. \emph{Biometrika} {\bf 57,} 97-109 (1970).

\bibitem[50]{Limousin_piemd} Limousin, M., Kneib, J.-P. \& Natarajan, P. Constraining the mass distribution of galaxies using galaxy-galaxy lensing in clusters and in the field. \emph{Mon. Not. R. Astron. Soc.} {\bf 356,} 309-322 (2005).

\bibitem[51]{Eliasdottir07} Eliasd\'ottir, \'A. \emph{et al.} Where is the matter in the Merging Cluster Abell 2218? \emph{arXiv e-prints,} arXiv:0710.5636 (2007).

\bibitem[52]{NFW96} Navarro, J. F., Frenk, C. S. \& White, S. D. M. The Structure of Cold Dark Matter Halos. \emph{Astrophys. J.} {\bf 462,} 563 (1996).

\bibitem[53]{Johnson17model} Johnson, T. L. \emph{et al.} Star Formation at z = 2.481 in the Lensed Galaxy SDSS J1110+6459. I. Lens Modeling and Source Reconstruction. \emph{Astrophys. J.} {\bf 843,} 78 (2017).

\bibitem[54]{Lejeune97} Lejeune, T., Cuisinier, F. \& Buser, R. Standard stellar library for evolutionary synthesis. I. Calibration of theoretical spectra. \emph{Astron. Astrophys. Suppl.} {\bf 125,} 229-246 (1997).

\bibitem[55]{Calzetti15} Calzetti, D. \emph{et al.} The Brightest Young Star Clusters in NGC 5253. \emph{Astrophys. J.} {\bf 811,} 75 (2015).

\bibitem[56]{Sanyal15} Sanyal,  D., Grassitelli, L., Langer, N. \& Bestenlehner, J. M. Massive main-sequence stars evolving at924the Eddington limit. \emph{Astron. Astrophys.} {\bf 580,} A20 (2015).

\bibitem[57]{ElBadry19_twinsies}  El-Badry, K., Rix, H.-W., Tian, H., Duch\^ene, G. \& Moe, M. Discovery of an equal-mass ‘twin’ binary population reaching 1000 + au separations. \emph{Mon. Not. R. Astron. Soc.} {\bf 489,} 5822-5857 (2019).

\bibitem[58]{Leitherer99} Leitherer, C. \emph{et al.} Starburst99: Synthesis Models for Galaxies with Active Star Formation. \emph{Astrophys. J. Suppl.} {\bf 123,} 3-40 (1999).

\bibitem[59]{Kroupa01_imf} Kroupa,  P.  On  the  variation  of  the  initial  mass  function. \emph{Mon. Not. R. Astron. Soc.} {\bf 322,} 231-246 (2001).

\bibitem[60]{daSilva12} da Silva, R. L., Fumagalli, M. \& Krumholz, M. SLUG—Stochastically Lighting Up Galaxies. I. Methods and Validating Tests. \emph{Astrophys. J.} {\bf 745,} 145 (2012).

\bibitem[61]{Krumholz15} Krumholz, M. R., Fumagalli, M., da Silva, R. L., Rendahl, T. \& Parra, J. SLUG -- stochastically lighting up galaxies - III. A suite of tools for simulated photometry, spectroscopy, and Bayesian inference with stochastic stellar populations. \emph{Mon. Not. R. Astron. Soc.} {\bf 452,} 1447-1467 (2015).

\bibitem[62]{MadauDickinson14} Madau, P. \& Dickinson, M. Cosmic Star-Formation History. \emph{Annu. Rev. Astron. Astrophys.} {\bf 52,} 415-486 (2014).

\bibitem[63]{ChabrierIMF}  Chabrier, G. Galactic Stellar and Substellar Initial Mass Function. \emph{PASP} {\bf 115,} 763-795 (2003).

\bibitem[64]{Kehrig18} Kehrig, C. \emph{et al.} The extended He II $\lambda$4686 emission in the extremely metal-poor galaxy SBS 0335 - 052E seen with MUSE. \emph{Mon. Not. R. Astron. Soc.} {\bf 480,} 1081-1095 (2018).

\bibitem[65]{Sarmento18} Sarmento, R., Scannapieco, E. \& Cohen, S. Following the Cosmic Evolution of Pristine Gas. II. The Search for Pop III-bright Galaxies. \emph{Astrophys. J.} {\bf 854,} 75 (2018).

\bibitem[66]{Sarmento19} Sarmento, R., Scannapieco, E. \& C\^ot\'e, B. . Following the Cosmic Evolution of Pristine Gas. III. The Observational Consequences of the Unknown Properties of Population III Stars. \emph{Astrophys. J.} {\bf 871,} 206 (2019).

\bibitem[67]{Windhorst18} Windhorst, R. A. \emph{et al.} On the Observability of Individual Population III Stars and Their Stellar-mass Black Hole Accretion Disks through Cluster Caustic Transits. \emph{Astrophys. J. Suppl.} {\bf 234,} 41 (2018).

\bibitem[68]{Dai-Pascale2021} Dai, L. \& Pascale, M.  New  Approximation  of  Magnification  Statistics  for  Random  Microlensing  of Magnified Sources. \emph{arXiv e-prints,} arXiv:2104.12009 (2021).

\bibitem[69]{yoli_whl_icl} Jim\'enez-Teja, Y. \emph{et al.} RELICS: ICL Analysis of the z = 0.566 merging cluster WHL J013719.8-08284. \emph{arXiv e-prints,} arXiv:2109.04485 (2021).

\bibitem[70]{Kriek09_fast} Kriek, M. \emph{et al.} An Ultra-Deep Near-Infrared Spectrum of a Compact Quiescent Galaxy at z = 2.2.\emph{Astrophys. J.} {\bf 700,} 221-231 (2009).

\bibitem[71]{Bruzual_Charlot_03} Bruzual, G. \& Charlot, S. Stellar population synthesis at the resolution of 2003. \emph{Mon. Not. R. Astron. Soc.} {\bf 344,} 1000-1028 (2003).

\bibitem[72]{Spera15} Spera, M., Mapelli, M. \& Bressan, A. he mass spectrum of compact remnants from the PARSEC stellar evolution tracks.

\bibitem[73]{Oguri2018} Oguri,  M.,  Diego,  J.  M.,  Kaiser,  N.,  Kelly,  P.  L.  \&  Broadhurst,  T.  Understanding  caustic  crossings in giant arcs: Characteristic scales, event rates, and constraints on compact dark matter. \emph{Phys. Rev. D} {\bf 97,} 023518 (2018).

\bibitem[74]{Trenti09} Trenti, M., Stiavelli, M. \& Shull, J. M. Metal-free Gas Supply at the Edge of Reionization: Late-epoch Population III Star Formation. \emph{Astrophys. J.} {\bf 700,} 1672-1679 (2009).

\bibitem[75]{Vanzella2020} Vanzella, E. \emph{et al.} Candidate Population III stellar complex at z = 6.629 in the MUSE Deep Lensed Field. \emph{Mon. Not. R. Astron. Soc.} {\bf 494,} L81-L85 (2020).

\bibitem[76]{ligo_190521} Abbott, R. \emph{et al.} GW190521: A Binary Black Hole Merger with a Total Mass of $150 M_{\odot}$. \emph{Phys. Rev. Lett.} {\bf 125,} 101102 (2020).

\bibitem[77]{farrell20_lowZbbh} Farrell, E. \emph{et al.} Is GW190521 the merger of black holes from the first stellar generations? \emph{Mon. Not. R. Astron. Soc. Lett.} {\bf 502,} L40-L44 (2020).

\bibitem[78]{kinugawa20_pop3bbh} Kinugawa, T., Nakamura, T. \& Nakano, H. Formation of binary black holes similar to GW190521 with a total mass of $150M_{\odot}$ from Population III binar star evolution. \emph{Mon. Not. R. Astron. Soc. Lett.} {\bf 501,} L49-L53 (2020).

\bibitem[79]{zdziarski-gierlinski04_bhxb} Zdziarski, A. A. \& Gierli\'nski, M. Radiative Processes, Spectral States and Variability of Black-Hole Binaries. \emph{Progress of Theoretical Physics Supplement} {\bf 155,} 99-119 (2004).

\bibitem[80]{Holwerda14} Holwerda, B. W. \emph{et al.} Milky Way Red Dwarfs in the BoRG Survey; Galactic Scale-height and the Distribution of Dwarf Stars in WFC3 Imaging. \emph{Astrophys. J.} {\bf 788,} 77 (2014).

\bibitem[81]{spex_browndwarf} Burgasser, A. J. \& Splat Development Team. The SpeX Prism Library Analysis Toolkit (SPLAT): A Data Curation Model. \emph{Astronomical Society of India Conference Series} {\bf 14,} 7-12 (2017).

\bibitem[82]{Hainline11_agn} Hainline, K. N., Shapley, A. E., Greene, J. E. \& Steidel, C. C. The Rest-frame Ultraviolet Spectra of UV-selected Active Galactic Nuclei at z $\sim$ 2-3. \emph{Astrophys. J.} {\bf 733,} 31 (2011).
  
\end{thebibliography}
\end{document}